%% file: jfm_RSS_revised.tex
\documentclass{jfm}

\usepackage{graphicx}
\usepackage{natbib}

\usepackage{color}
\usepackage{cancel}
\usepackage[normalem]{ulem}
\ifCUPmtlplainloaded \else
  \checkfont{eurm10}
  \iffontfound
    \IfFileExists{upmath.sty}
      {\typeout{^^JFound AMS Euler Roman fonts on the system,
                   using the 'upmath' package.^^J}%
       \usepackage{upmath}}
      {\typeout{^^JFound AMS Euler Roman fonts on the system, but you
                   dont seem to have the}%
       \typeout{'upmath' package installed. JFM.cls can take advantage
                 of these fonts,^^Jif you use 'upmath' package.^^J}%
      }
  \else
  \fi
\fi

\ifCUPmtlplainloaded \else
  \checkfont{msam10}
  \iffontfound
    \IfFileExists{amssymb.sty}
      {\typeout{^^JFound AMS Symbol fonts on the system, using the
                'amssymb' package.^^J}%
       \usepackage{amssymb}%

      }{}
  \fi
\fi

\ifCUPmtlplainloaded \else
  \IfFileExists{amsbsy.sty}
    {\typeout{^^JFound the 'amsbsy' package on the system, using it.^^J}%
     \usepackage{amsbsy}}
    {}
\fi

\newsavebox{\astrutbox}
\sbox{\astrutbox}{\rule[-5pt]{0pt}{20pt}}

\newcommand{\ds}[2]{\displaystyle{\frac{#1}{#2}}}
\newcommand{\bfu}{\mathbf u}
\newcommand{\bfn}{\mathbf n}
\newcommand{\bfta}{{\mathbf t}_1}
\newcommand{\bftb}{{\mathbf t}_2}
\newcommand{\bfT}{\mathbf T}
\newcommand{\bfI}{\mathbf I}
\newcommand{\bfD}{\mathbf D}
\newcommand{\bfe}{\mathbf e}
\newcommand{\g}{\nabla}
\newcommand{\cH}{{\cal H}}
\newcommand{\cP}{{\cal P}}
\usepackage{bm}
\usepackage{tikz}
\usetikzlibrary{arrows}
\def\bk{\bm{k}}

\newcommand{\acs}[1]{{{#1}}}
\renewcommand{\sout}[1]{}
\renewcommand{\cancel}[1]{}

\title[Faraday waves]{Can weakly nonlinear theory explain Faraday wave
patterns near onset?}

\author[A. C. Skeldon and A. M. Rucklidge]%
{A.\ns C.\ns S\ls K\ls E\ls L\ls D\ls O\ls N$^1$
  \thanks{Email address for correspondence: a.skeldon@surrey.ac.uk},\ns
and A.\ns M.\ns R\ls U\ls C\ls K\ls L\ls I\ls D\ls G\ls E$^2$%
}

\affiliation{$^1$Department of Mathematics, University of Surrey, 
 Guildford, Surrey, GU2 7XH, UK \\[\affilskip]
$^2$Department of Mathematics, University of Leeds,
Leeds, LS2 9JT, UK
 }

\pubyear{}
\volume{}
\pagerange{}
\date{?; revised ?; accepted ?. - To be entered by editorial office}
\begin{document}

\maketitle
\begin{abstract}
The Faraday problem is an important pattern-forming system that 
provides some middle ground between systems where the initial 
instability involves just 
a single mode and in which complexity then results from mode interactions 
or secondary bifurcations, 
and cases where a system is highly turbulent
and many spatial and temporal modes are excited.  It has
been a rich source of novel patterns and of theoretical work aimed at understanding
how and why such patterns occur.  Yet it is particularly
challenging to tie theory to experiment:  the experiments are difficult to perform;
the parameter regime of interest (large box, moderate viscosity) along
with the technical difficulties of solving the free boundary Navier--Stokes equations
make numerical solution of the problem hard; and the fact that the instabilities 
result in an entire circle of unstable wavevectors presents considerable 
theoretical difficulties.
 
In principle, weakly nonlinear theory should be able to predict which patterns
are stable near pattern onset.  
In this paper we present the first quantitative comparison between 
weakly nonlinear theory of the full Navier--Stokes equations and
(previously published) experimental results for the 
Faraday problem with multiple frequency forcing. 
We confirm that three-wave interactions sit at the heart of why complex
patterns are stablised but also highlight some discrepancies between theory and
experiment. These suggest the need for further experimental and theoretical
work to fully investigate the issues of pattern bistability and the role
of bicritical/tricritical points in determining bifurcation structure.

\end{abstract}


\section{Introduction}

Since \cite{F_1831} identified that regular patterns can appear on
the surface of a shaken container of fluid, 
the Faraday experiment has been an important system
for investigating pattern formation.  Experiments in the 1980's such as those 
conducted
by \cite{SG_JFM89} tended to focus on the dynamics of the interaction of patterns in
small containers excited by a sinusoidal forcing with a single frequency 
component. More recently the focus has switched to larger containers with
multiple frequency components:  
as pointed out in \cite{AF_PRL98}, one special feature of the 
Faraday experiment is that by using multiple frequency forcing one
can investigate the interaction of a small number of controllable modes
with different characteristic length scales.  This 
provides some middle ground between pattern-forming systems where the initial 
instability involves just 
a single mode and in which complexity then results from secondary bifurcations, 
such as in the 
B{\'e}nard--Marangoni experiment or the Taylor-Couette experiment, 
and cases where a system is highly turbulent
and many spatial and temporal modes are excited.



Within the Faraday experiment, a rich variety of complex patterns
are seen, some of which have a complicated spatial structure but are 
time-periodic with
the periodicity of the drive and some with both a complicated spatial and
temporal structure.  Aside from the ubiquitous stripes, squares and hexagons,
observed patterns include: 
quasipatterns (\cite{Christiansenetal_PRL92}, \cite{EF_JFM94});
superlattice patterns (SL1) (\cite{KPG_PhysD98,EF_PRE06}); 
spatially subharmonic superlattice states, 
modulated hexagonal disorder, two mode 
superlattices and unlocked states (\cite{AF_PRL98});  oscillons 
(\cite{AF_PRL00b}), and double hexagon states \cite{AF_PRL98}.  An excellent summary
of many of the experimental results is given in  \cite{AF_PRE02}.

In parallel with the experiments, theoretical advances have resulted in
an effective numerical method for performing the linear stability analysis 
of the Navier--Stokes equation that marks
the transition from an unpatterned to patterned state 
(\cite{KT_JFM94}).  The challenge here is that many of the experiments 
are carried out
at moderate viscosity whereas early theoretical results of \cite{BU_PROC54},
which help explain the underlying instability mechanism, are for an 
inviscid fluid. There have also been 
theoretical explanations for why, unlike the 
B{\'e}nard--Marangoni system, the observed patterned states include superlattice
patterns (\cite{SS_PRE99,STS_PhysD00,SP_PRL98, RucklidgeSilber_SIADS09}).

These theoretical mechanism\acs{s} rely on three-wave resonance between critical
modes and modes close to critical and have successfully explained why
particular superlattice patterns are observed.
The suggested theoretical mechanism for the appearance of superlattice 
patterns is compelling, and the link between which modes are excited and
which patterns are observed has been explored in some detail experimentally,
for example see \cite{AF_PRE02,EF_PRE06}.
However, without carrying
out a careful quantitative comparison between experiment and
theory, it is hard to know the extent to which the theoretical ideas 
really do explain the 
experimental results (\cite{BPA_AnnRevFM00}).  
This is particularly true for the Faraday 
experiment where, as we will explain in greater detail below, the very 
region for which the theory predicts superlattice patterns is the region for which 
some of the underlying assumptions of the theory break down.

While it is not possible to write down closed-form solutions of the full
Navier--Stokes equations for the Faraday problem, quantitative predictions of 
the patterns expected near the transition from non-patterned to pattern states
can be made using weakly nonlinear analysis (\cite{SG_SIAP07}).  
Weakly nonlinear analysis centres
on using an asymptotic expansion in terms of the slowly varying amplitudes
of the critical modes at onset, using ideas first developed in \cite{MV_JFM58}
and \cite{SS_JFM62} in the context of convection experiments.  
Analysis of the resulting amplitude equations
leads to predictions on the relative stability of different
patterned states.

Our aim in this paper is to do a quantitative comparison between 
previously published experimental results
and a weakly nonlinear analysis of the Navier--Stokes equations
for the Faraday problem, \acs{with a particular focus on multiple frequency
forcing}.
\acs{A previous comparison between the theory and experiments in the 
Faraday problem was carried out by \cite{WBW_JFM03} for single frequency
forcing in which the authors declare that the excellent agreement they 
observe means that the Faraday problem is essentially a `solved problem'.  
In particular, their paper uses a Lyapunov stability argument based
on the theory of \cite{CV_PRE99} to find
the `most stable' pattern as a function of the key non-dimensional groups
in the problem representing non-dimensional measures of viscous, gravitational
and surface tension forces. Results from this 
theoretical study agree very well with the author's experimental results. 
However, 
the extension to multiple frequency forcing is far from trivial.  The
method used by \cite{CV_PRE99} to reduce the Navier-Stokes equations to
amplitude equations is not applicable, and the more general theory in
\cite{SG_SIAP07} is needed; the addition of more frequency components
introduces more non-dimensional parameters, resulting in a much greater ability
to probe underlying three-wave mechanisms.  Consequently, patterns such
as superlattice patterns, 
which are not observed in the single frequency context, are found. In
fact, as we will show, for multiple frequency excitation there remain
many open questions.   
We note also that \cite{WBW_JFM03} use a Lyapunov stability argument 
to determine preferred patterns.  We use similar arguments, but
have in addition carried out
a bifurcation analysis of the relevant amplitude equations.  This has the
added benefit of not only determining the `most stable' pattern but also 
indicating regions where patterns are bistable.

Specifically, in this paper we
\sout{We will}} discuss to what extent the existing weakly nonlinear theory
can explain observed patterns \acs{in multiple-frequency Faraday experiments} 
and provide 
some new explanations in some cases.  While agreement is very good in many
cases, we note that quantitatively linking theory with experiment is particularly
challenging after-the-fact as the results are sensitive to precise values
of viscosity and surface tension and even the sign of the drive term, something
that is not normally recorded.  There are also places where the 
analysis strongly suggest\acs{s} that patterns should have a subharmonic component, when
no subharmonic component has been observed.  In particular, with regard to
the superlattice patterns, we discuss two methods that have been
used to promote the stability of superlattice patterns: firstly
by approaching the so-called bicritical point in two-frequency forced
experiments; and, secondly, by adding a third frequency to the drive. 
We highlight the differences between these two mechanisms. 

Overall,
we confirm that three-wave interactions sit at the heart of why complex
patterns are stablised. However, the discrepancies between theory and
experiment suggest the need for further experimental and theoretical
work to fully investigate the issues of pattern bistability and the role
of bicritical/tricritical points in determining bifurcation structure.

\section{Equations}
Using a variety of container shapes, 
\cite{EF_JFM94} elegantly demonstrated 
that many of the patterns that occur with
moderate viscosity fluids in large containers are not
strongly dependent on the lateral boundaries of the container.  Consequently
it is a reasonable modelling assumption to 
consider an infinite horizontal layer of viscous incompressible fluid
of finite depth that is subjected to gravity \acs{$\tilde g$} and to a vertical periodic
excitation with \acs{non-dimensional} frequency components $j \acs{\omega}$. 
At the lower boundary the fluid is in
contact with a rigid plane while at the upper boundary the surface
is open to the atmosphere.
This means that the upper surface is a free boundary whose shape
and evolution is an unknown of the problem.

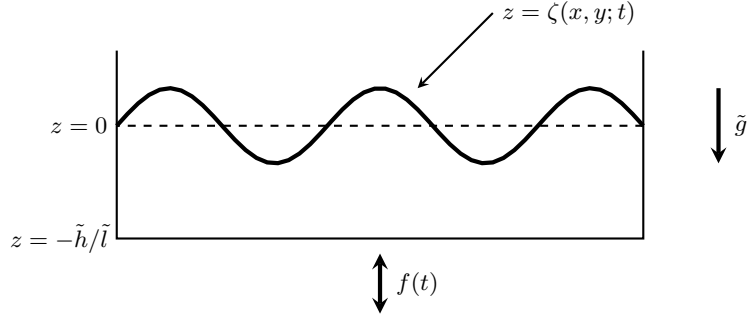
\begin{figure}
\begin{center}
\input{fig1.tex}
\end{center}
\caption{Sketch of a cross-section through the layer of fluid.}
\label{geom}
\end{figure}

The motion of the fluid can be described by the Navier--Stokes equations,
where, to take account of the parametric excitation, a 
 frame of reference which is moving with the
periodic excitation is considered. 
The $z$-axis is chosen perpendicular to the rigid
plane at the bottom, which lies at $z=-\acs{\tilde h/\tilde l}$, where 
$\acs{\tilde h/\tilde l}$ 
is the non-dimensional
depth of the layer when the fluid is at rest.  
A sketch of the geometry is shown in figure~\ref{geom}.
Assuming the free surface may be written as $z=\zeta(x,y;t)$, which excludes
the formation of droplets or breaking waves, 
then the fluid motion in the bulk is described by the 
dimensionless incompressible
Navier--Stokes equations
\begin{eqnarray}
\g \cdot \bfu & = & 0, \nonumber \\
\partial_t \bfu + \bfu \cdot \g \bfu & = & - \g \cP + C
\Delta \bfu - (1+ f(t) )\bfe_3, \label{NSnd}
\end{eqnarray}
where $\bfu=(u,v,w)$ is the velocity field, $\cP$ the pressure and
for multiple frequency excitation,
\begin{equation}
f(t)= \sum_j \acs{a_j} \cos(j \acs{\omega} t + \phi_j),
\label{drive}
\end{equation}
where $j$ are integers
and \acs{the non-dimensional amplitudes} $\acs{a_j}$ and \acs{phases} $\phi_j$ are real. 

It is assumed that the bottom of the container
is rigid so that at $z=-\acs{\tilde h/\tilde l}$ the
fluid satisfies the no-slip boundary
conditions 
\begin{equation}
u=v=w=0. \label{bctop}
\end{equation}
At the free surface $z=\zeta(x,y;t)$ we have the kinematic
condition,  which says that the surface is advected by the fluid, and
two further conditions, one for the balance of the tangential stresses
and one for the balance of normal
stresses. This leads to three conditions at
$z=\zeta(x,y;t)$ namely
\begin{eqnarray}
\partial_t \zeta + u \partial_x \zeta + v
\partial_y \zeta & = &  w, \nonumber \\
\bfta \cdot \bfT \bfn = \bftb\cdot \bfT \bfn & = & 0, \label{bcbottom} \\
-\cP+2 C \bfn \bfD(\bfu) \bfn & =  & B \cH - p_e, \nonumber
\end{eqnarray}
where $\bfT=-\cP \bfI +2 C \bfD(\bfu)$ is the stress tensor,
$\bfD(\bfu)=(\g \bfu + \g^T \bfu)/2$ is the rate-of-strain tensor,
and $\cH = \g_H \cdot(\g_H \zeta / \sqrt{1+|\g_H \zeta|^2})$
is the double mean curvature. Note that
$\nabla=(\g_H,\partial_z)$ and $\g_H=(\partial_x,\partial_y)$. 
The unit normal and
tangent vectors are defined as
\begin{eqnarray*}
\bfn(x,y;t) & = & \left(-\ds{\partial_x \zeta}{\sqrt{1+|\g_H
\zeta|^2}}, -\ds{\partial_y \zeta}{\sqrt{1+|\g_H \zeta|^2}},
\ds{1}{\sqrt{1+|\g_H \zeta|^2}}\right), \\
\bfta(x,y;t)& = & \left(\ds{1}{\sqrt{1+|\partial_x \zeta|^2}},0,
\ds{\partial_x \zeta}{\sqrt{1+|\partial_x \zeta|^2}}\right),  \\
\bftb(x,y;t) & = & \left(0, \ds{1}{\sqrt{1+|\partial_y \zeta|^2}},
\ds{\partial_y \zeta}{\sqrt{1+|\partial_y \zeta|^2}}\right),
\end{eqnarray*}
The units of length, time, velocity
and pressure have been taken as $\acs{\tilde l}, \sqrt{\acs{\tilde l/\tilde g}}, 
\sqrt{\acs{\tilde g \tilde l}}$ and
$\acs{\tilde \varrho \tilde g \tilde l}$ respectively, 
\acs{where, $\tilde \varrho$ is the density of the fluid and $\tilde g$ is
the acceleration due to gravity}.
The length scale $\acs{\tilde l}$ is taken to be a length
scale that is typical for the problem, such as the primary
wavelength of the observed patterns.
Here, $p_e$ is the dimensionless pressure of the external
ambient fluid and is assumed known. 
There are two non-dimensional parameters associated with
the fluid, namely: $C = \acs{\tilde \nu/(\tilde g \tilde l^3})^{1/2}$, 
the square of the inverse of the Galileo
number, 
and $B=\acs{\tilde \sigma/\tilde \varrho \tilde g \tilde l^2}$, the inverse Bond number,
where $\acs{\tilde \nu}$ is the kinematic viscosity \acs{and}
$\acs{\tilde \sigma}$ is the surface tension. 
These two non-dimensional parameters
measure the relative importance of 
viscous and surface tension forces compared to gravity respectively.
There are three other sets of nondimensional parameters of importance,
all associated with the excitation.  These are:
the non-dimensional amplitudes $\acs{a_j}$, frequencies $j \acs{\omega}$  
and phases $\phi_j$ of the 
components of the excitation.  \acs{In the comparison with experiments
it is frequently useful to quote the relevant
dimensional values: we have used the convention that all dimensional variables
are labelled with a tilde.}

As in \cite{KT_JFM94}, it is convenient to define a new pressure,
\begin{equation}
p=\cP+ (1+ f(t))z,
\end{equation}
which has the effect of shifting the acceleration term from the
momentum equation to the normal stress condition. In addition, we
eliminate the pressure from the momentum equation by taking $-(\g
\times\g \times)$. Using the relation $\g \times \g \times \bfu
=\g(\g \cdot \bfu )-\Delta \bfu$ and the fact that $\g \cdot
\bfu=0$, the problem then becomes
\begin{eqnarray}
\g \cdot \bfu & = & 0, \nonumber \\
\partial_t \Delta \bfu -C \Delta \Delta \bfu & = &
\g \times\g \times( \bfu \cdot \g \bfu ),
\label{NSndf}
\end{eqnarray}
with boundary conditions on $z=-\acs{\tilde h/\tilde l}$,
\begin{equation}
u=v=w=0 \qquad \label{bctopp}
\end{equation}
and on $z=\zeta$,
\begin{eqnarray}
\partial_t \zeta + u \partial_x \zeta + v
\partial_y \zeta & = & w, \nonumber \\
\bfta \cdot \bfT \bfn = \bftb\cdot \bfT \bfn & = & 0, \label{bcbottomp} \\
2 C \bfn \bfD(\bfu) \bfn & = &
B \cH  + p - p_e - (1+ f(t)) \zeta.
\nonumber
\end{eqnarray}

Equations (\ref{NSndf}) with boundary conditions (\ref{bctopp})
and (\ref{bcbottomp}) have a trivial solution,
\begin{equation}
\bfu = {\bf0}, \quad p=p_e, \quad \zeta =0.
\label{eq:trivialsol}
\end{equation}
This solution corresponds to an unpatterned state where there is
no relative motion of the fluid with respect to the moving frame so the
surface of the fluid is flat.

\section{Does linear theory agree with experiments?}

Understanding the linear stability of the unpatterned state is
at the heart of understanding many of the nonlinear patterns that are formed
close to onset. This is because 
when the unpatterned state becomes unstable to one critical mode with
a given critical wavenumber, there
are often several other modes with different wavenumbers
that are themselves only weakly damped. This is particularly true with 
multi-frequency forcing.
Resonant interaction of 
instabilities from different critical/close to critical modes drive
the selection mechanisms for the occurrence of particular patterns 
(\cite{STS_PhysD00}). 

\cite{BU_PROC54} recognised that the linear stability of the flat-surface
solution for an inviscid, infinite depth fluid driven by a single
frequency reduces to a Mathieu equation.  
The Mathieu equation contains two parameters, related to the 
frequency
$\tilde \omega$
and amplitude $\tilde a$ of the excitation respectively.  Solutions to the 
Mathieu
equation divide the parameter plane into regions of bounded and regions of
unbounded solutions where the regions of unbounded solutions form tongues
that touch the frequency axis at frequencies $m \tilde \omega/2, m=1,2...$, 
the largest tongue occuring for 
$m=1$.  The tongues are typically
classified as either harmonic or subharmonic, depending on whether
or not they are an integer multiple of the 
frequency $\tilde \omega$.
This picture is modified with the addition of 
damping: the boundaries of the regions are perturbed and no longer
touch the frequency axis, consequently a finite amplitude of excitation
is required to excite waves;
the regions of bounded solutions become regions where the
unpatterned state 
is locally stable; the unbounded regions become regions where the
unpatterned state is unstable. 

\begin{figure}
\begin{center}
\includegraphics[scale=1]{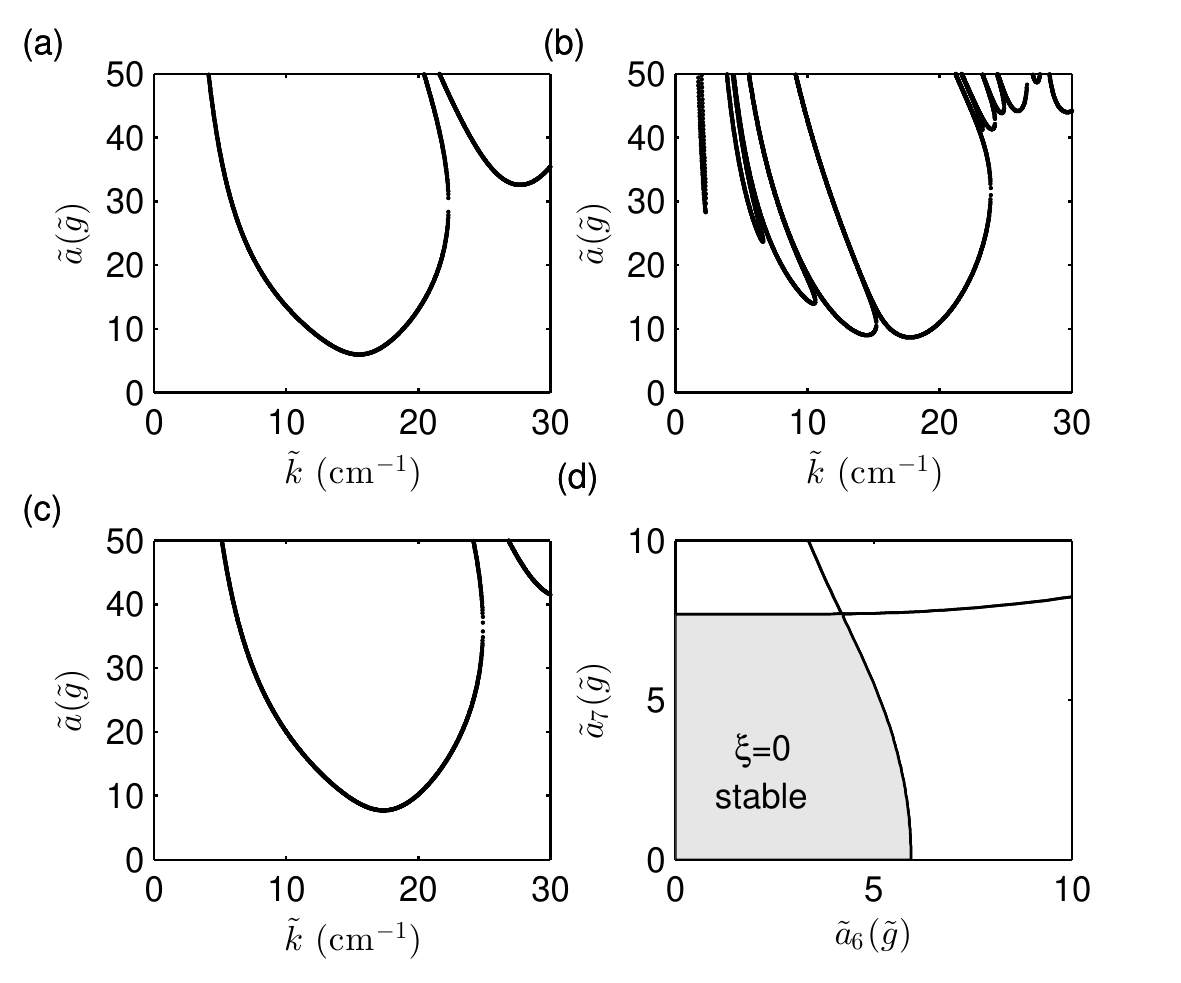}
\end{center}
\caption{Typical tongues and linear stability region for 
the unpatterned state computed using the method of 
\cite{BET_PRE96}.  Parameters as for 
\cite{KPG_PhysD98}: the (dimensional) excitation is 
$\acs{\tilde f(\tilde t)} 
= \acs{\tilde a_6} \cos 6 \acs{\tilde \omega} \acs{\tilde t} 
+ \acs{\tilde a_7} \cos (7 \acs{\tilde \omega} \acs{\tilde t} + \phi_7)$, 
where $\acs{\tilde a_6} = \acs{\tilde a} \cos \chi$,
$\acs{\tilde a_7} = \acs{\tilde a} \sin \chi$, $\phi_7=20^\circ$, 
$\acs{\tilde \omega}/2 \pi = 16.44$ Hz, 
$\acs{\tilde \sigma}=20.6$ dyn cm$^{-1}$,
$\acs{\tilde \nu}=0.20$ cm$^2$s$^{-1}$,
$\acs{\tilde \varrho}=0.95$ g cm$^{-3}$,
$\acs{\tilde h}=0.3$ cm. 
Note that the structure inherited from the Mathieu equation is seen by the
fact that the minimum of each tongue occurs at half integer multiples
of the drive frequency. 
In (a) $\chi =0^\circ$, the forcing reduces to 
$\acs{\tilde f(\tilde t)}= \acs{\tilde a_6} \cos 6 \acs{\tilde \omega} \acs{\tilde t}$,
with a forcing frequency of $6 \acs{\tilde \omega}$.  
The largest tongue
has a minimum for wavenumber of approx $15$cm$^{-1}$ for a forcing
amplitude $\acs{\tilde a}$ of approx $7.5\acs{\tilde g}$ and corresponds to a mode of frequency   
$\frac{6 \acs{\tilde \omega}}{2}$. In  (c), $\chi=90^\circ$ the forcing reduces
to $\acs{\tilde f(\tilde t)}=\acs{\tilde a_7} \cos 7 \acs{\tilde \omega} \acs{\tilde t}$ 
with a forcing frequency of $7 \acs{\tilde \omega}$
leading to a primary instability mode with
frequency $7 \acs{\tilde \omega}/2$. 
In the case of (b), $\chi=63^\circ$, the presence of both $6 \acs{\tilde \omega}$
and $7 \acs{\tilde \omega}$ components in the forcing frequency mean that the
drive has periodicity $2 \pi/\acs{\tilde \omega}$ and is therefore of 
frequency $\acs{\tilde \omega}$. This leads to tongues at 
$m\acs{\tilde \omega}/2\ldots, m=1,2,\ldots$. 
The first five tongues that are visible from left to right 
correspond to: $\acs{\tilde \omega}/2, 4\acs{\tilde \omega}/2, 
5 \acs{\tilde \omega}/2, 6 \acs{\tilde \omega}/2, 7 \acs{\tilde \omega}/2$.  
The largest tongues correspond to the
$6\acs{\tilde \omega}/2$ and the $7 \acs{\tilde \omega}/2$ modes and
are driven by the two main frequency components
of the drive.  There are tongues corresponding to modes with 
frequency $2\acs{\tilde \omega}/2$ and $3\acs{\tilde \omega}/2$ 
but these occur off the top of the
region shown.
This particular value of $\chi$ is close to the `bi-critical' point
where both harmonic modes with frequency $6 \acs{\tilde \omega} /2$ and sub-harmonic modes
with frequency $7 \acs{\tilde \omega}/2$ onset simultaneously.  It is close to this bi-critical
point that many of the exotic patterns are observed.
(d) Bifurcation set showing the position of the tongue minimum that marks the instability
of the unpatterned state, as a function of $\acs{\tilde a_6}$ and $\acs{\tilde a_7}$.  } 
\label{fig:tongues}
\end{figure}

\cite{KT_JFM94} identified a numerical method to find 
the instability tongues
that can be used for all fluid viscosities and all depths and applied it
to the case of single frequency excitation. This was extended to 
multiple frequency excitation
by \cite{BET_PRE96}. 
A typical example of tongues
computed using \cite{BET_PRE96}'s method and the corresponding bifurcation
set for the primary stability boundary is shown in figure \ref{fig:tongues}.

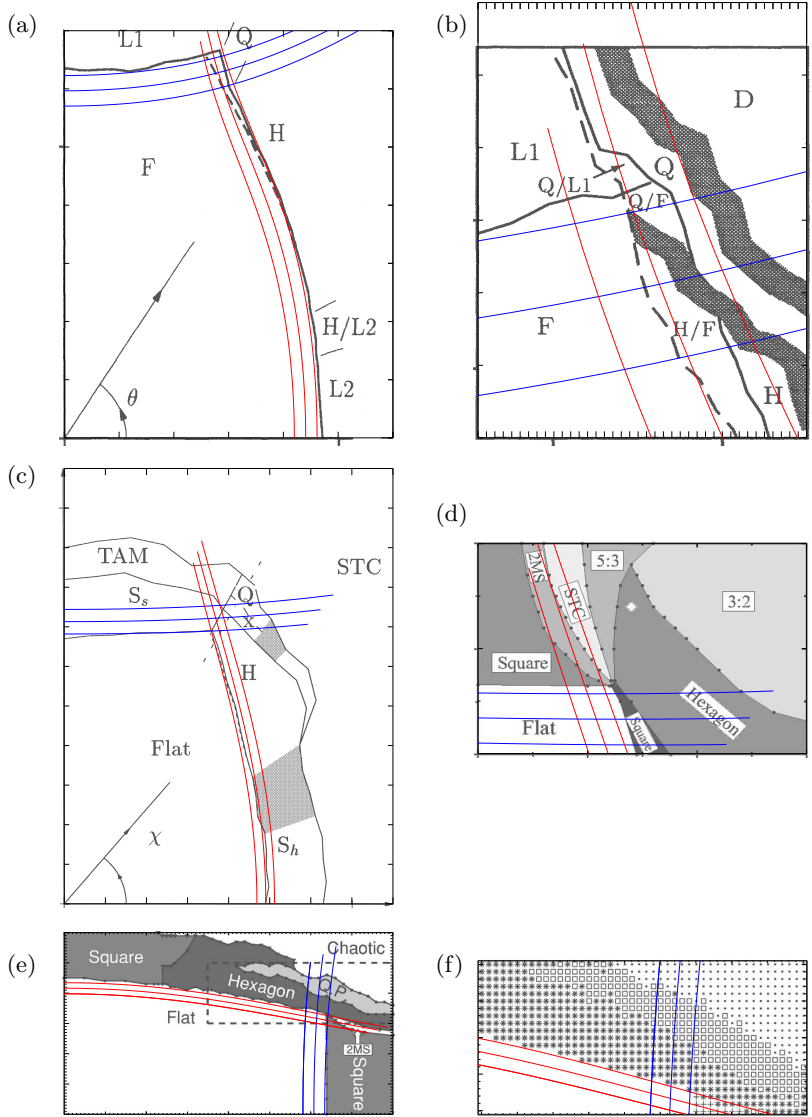
\begin{figure}
\input{fig3.tex}
%
%
%
%
\caption{
Comparison of the experimentally measured transition from unpatterned to patterned
state with numerically calculated linear stability curves using the method
of \cite{BET_PRE96}. For each of the numerical calculations, 
three different values for the viscosity are used.
(a) and (b) 
Numerical curves overlaid on figure 12 of \cite{EF_JFM94}.
Adapted with permission \acs{from \cite{EF_JFM94}}.
$\acs{\tilde \sigma}=65$ dyn cm$^{-1}$; 
$\acs{\tilde \nu}=1.00\pm0.05$ cm$^2$ s$^{-1}$; 
$\acs{\tilde \varrho}=1.22$ g m$^{-3}$;
$\acs{\tilde h}=0.29$ cm; 
$\acs{\tilde a_j}, j=\{4,5\}$; 
$\phi_4=0$;
$\phi_5=75^\circ$;
$\acs{\tilde \omega}/2\pi=14.6$ Hz.
(c) Numerical curves overlaid on figure 6(a)
of \cite{KPG_PhysD98}
reprinted from Physica D, Vol 123,
A. Kudrolli, B. Pier and J.P. Gollub,
`Superlattice patterns
in surface waves', p99--111, 
Copyright (1998), with permission from
Elsevier.
$\acs{\tilde \sigma}=20.6$ dyn cm$^{-1}$; 
$\acs{\tilde \nu}=0.20 \pm 0.01$ cm$^2$ s$^{-1}$; 
$\acs{\tilde \varrho}=0.95$ g m$^{-3}$; 
$\acs{\tilde h}=0.3$ cm;
$\acs{\tilde a_j}, j=\{4,5\}$; 
$\phi_4=0$;
$\phi_5=16^\circ$;
$\acs{\tilde \omega}/2\pi=22$ Hz.
(d) Numerical curves overlaid on figure 2 of \cite{EF_PRE06}. 
Adapted with permission from \cite{EF_PRE06}.  Copyrighted by
the American Physical Society.
$\acs{\tilde \sigma}=20.6$ dyn cm$^{-1}$; 
$\acs{\tilde \nu}=0.180\pm 0.009$ cm$^2$ s$^{-1}$; 
$\acs{\tilde \varrho}=0.949$ g m$^{-3}$; 
$\acs{\tilde h}=0.3$ cm;
$\acs{\tilde a_j}, j=\{6,7\}$; 
$\phi_j=0^\circ$;
$\acs{\tilde \omega}/2\pi=14$ Hz.
(e) and (f) Numerical curves overlaid on figures 1 and
3(a) of \cite{DU_PRE06}. 
Adapted with permission from \cite{DU_PRE06}.  Copyrighted by
the American Physical Society.
$\acs{\tilde \sigma}=20.6$ dyn cm$^{-1}$; 
$\acs{\tilde \nu}=0.20 \pm 0.01$ cm$^2$ s$^{-1}$; 
$\acs{\tilde \varrho}=0.95$ g m$^{-3}$; 
$\acs{\tilde h}=0.65$ cm;
$\acs{\tilde a_j}, j=\{4,5\}$; 
$\phi_4=0$;
$\phi_5=16^\circ$;
$\acs{\tilde \omega}/2\pi=20$ Hz.
}
\label{fig:linearstabilitycomparison}
\end{figure}

The linear stability problem appears to be solved: the numerical
method works well and in \cite{BET_PRE96} 
the authors show that there is 
excellent agreement between their linear stability calculations
and experiments. 
However, unless experiments and theory are
carried out hand-in-hand, in practice there remain some 
difficulties in obtaining really good agreement between theory and experiment 
even for
this first transition from patterned to unpatterned state.  This is illustrated
in figure~\ref{fig:linearstabilitycomparison} where numerical linear 
stability results are
superimposed on the experimental bifurcation sets 
published in: \cite{EF_JFM94} (panels (a) and (b)); 
\cite{KPG_PhysD98} (panel (c)); \cite{EF_PRE06} (panel (d)), and  
\cite{DU_PRE06} (panels (e) and (f)).  

The central issue here is that the position of the curves 
is sensitive to the values of the surface tension, density and the viscosity and
yet the values quoted in papers are often taken from the manufacturers'
specifications for the fluids used. In order to illustrate the issue, in
each case we have plotted the linear 
stability curves for the quoted viscosity and the quoted viscosity plus or
minus 5\%, a typical quoted tolerance for the viscosity value.  
From figure~\ref{fig:linearstabilitycomparison} we see that: for the
experiments in \cite{EF_JFM94} and \cite{DU_PRE06}, 
the upper extreme for the viscosity fits the
data best (panels (a),(b),(e) and (f)); 
for those in \cite{KPG_PhysD98}, the lower extreme fits the subharmonic
boundary best, but not the harmonic boundary (panel (c)); 
for the experiments in \cite{EF_PRE06}
the linear stability analysis suggests that the actual viscosity of the fluid
was higher than the upper value (panel (d)).  We have focussed here on viscosity because for the different
fluids used in the results presented here, errors in the viscosity have the biggest
effect on the linear stability boundary.  A\acs{n} error in the viscosity of 5\% can result
in an error in the linear stability boundary of 5\%, whereas a
5\% error in either the surface tension or the density leading to a
5\% error in the ratio $\acs{\tilde \sigma/\tilde \varrho}$ 
leads to an error in the linear stability
boundary of only around 1\%.  

Our aim in this paper is to compare weakly nonlinear theory with experiments. 
Since the transitions from one preferred patterns to another turn out to
be quite delicate, clearly
it is only realistic to hope for agreement for nonlinear pattern selection
if there is first excellent agreement with linear theory.  
In making our nonlinear comparisons below, we have therefore used values of the
viscosity found by fitting the linear theory to the published data.

\section{Predicting patterns close to onset} 
\subsection{Theoretical ideas}\label{subsect:theoreticalideas}
Once a mode with a given wave number has become unstable, the 
fact that there is no preferred horizontal direction means that
standing waves of any orientation can occur.  In the 
B{\'e}nard--Marangoni
experiment,
which has a similar orientational invariance, this means that
patterns consisting of nonlinear superpositions of modes with
wave vectors of the same wave number but different orientations often 
lead to the observation of patterns with, for example, hexagonal symmetry.  
In principle, superlattice patterns could also
occur in convection experiments (\cite{SS_PhysD98}), but in practice they have 
not been seen without the addition of a vertical oscillation, as
in \cite{RSBP_PRL00}.   

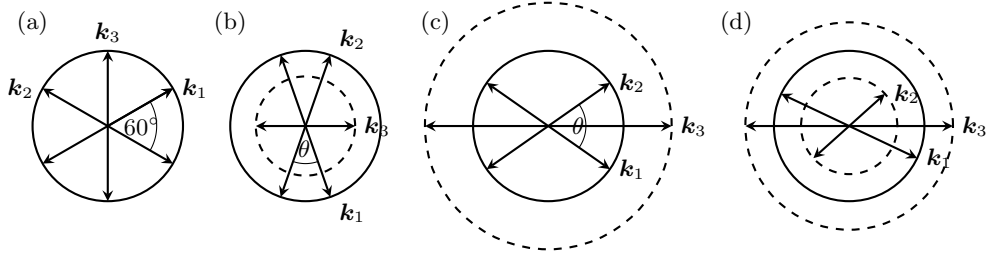
\begin{figure}
\input{fig4}
\caption{Possible spatial three-wave resonant interactions
defined such that $\bf k_1+k_2=k_3$.
(a) $\bf |k_1|=|k_2|=|k_3|$;
(b) $\bf |k_1|=|k_2|>|k_3|$;
(c) $\bf |k_1|=|k_2|<|k_3|$;
(d) $\bf |k_1|\neq|k_2|\neq|k_3|$.
}
\label{fig:threewave}
\end{figure}

The ubiquitous occurrence of hexagons is a consequence of the 
importance of three-wave interactions in determining which patterned states
occur:  three-wave interactions 
give rise to the lowest order nonlinear (quadratic) terms in the 
amplitude equations that describe behaviour close to onset.  
For B{\'e}nard--Marangoni convection, where instability to a 
single dominant wavenumber occurs, the three-wave interaction
of importance occurs between three wave vectors with the same critical
wavenumber, as shown in figure~\ref{fig:threewave}(a).    A distinct
feature of the Faraday problem is that the Mathieu tongue-like
structure to the linear stability problem means that although 
typically there
is  a single mode that becomes unstable first, there 
are nearby modes with different wavenumbers 
that are only weakly damped.  This can give rise
to other three-wave interactions of relevance, such as those shown in
figure~\ref{fig:threewave}(b), (c) and (d).

These three-wave interactions
involve waves with wavevectors $\bf k_1, k_2$ and $\bf k_3$ and
respective frequencies $n\omega/2, p\omega/2$ and $q \omega$.  
For the 
three waves to interact they must satisfy a spatial 
resonance condition, 
\begin{equation}
\bf \pm k_1 \pm k_2 = k_3, 
\label{eq:spatialresonancecondition}
\end{equation}
and a temporal resonance condition. \acs{In the spatial resonance
condition the choice of sign arises because the waves are standing waves and
so have spatial fourier components with both signs of wave vector}.
The temporal resonance condition
depends on both the frequencies of the waves and the various
frequencies contained within the forcing term $f(t)$ and
requires
\begin{equation}
\pm \frac{n}{2} \pm \frac{p}{2} \pm \acs{\cancel{\sum_j}} j = q
\label{eq:temporalresonance}
\end{equation}
where $j$ \acs{\sout{come from the forcing} 
 is one of
the frequency} component\acs{s} 
of the drive 
see equation (\ref{drive}),\acs{cf \cite{TPS_PRE04}} and
where all possible sign combinations of the different terms on the
left-hand side need to be considered \acs{because the waves are standing
waves and the forcing is real}.  

The fact that a temporal resonance condition needs to be satisfied
is one feature that distinguishes the Faraday problem from 
Swift--Hohenberg multiple resonance problems such as those studied
in \cite{Muller_PRE94}, \cite{LP_PRL97}, \cite{RSS_PRL12} and others.  This means that 
results from Swift--Hohenberg-like
equations need to be interpreted with care when applied to the Faraday
problem, a point that we will return to in the discussion.  

For the appearance of superlattice patterns 
the argument goes that, given three critical modes ${\bf k}_i$ as shown in 
figure~\ref{fig:threewave}(b) or (c) with amplitudes $A_i$, 
then using a multiple timescale expansion near onset
would lead to equations for the evolution of the amplitudes on a slow 
timescale of the form
\begin{eqnarray}
\dot A_1 & = & \lambda_1 A_1 + \alpha_1 \overline{A_2} A_3 + 
A_1 ( a |A_1|^2 + b_0 |A_2|^2 + c |A_3|^2 ) + \ldots \nonumber \\
\dot A_2 & = & \lambda_1 A_2 + \alpha_1 \overline{A_1} A_3 + 
A_2 ( b_0 |A_1|^2 + a |A_2|^2 + c |A_3|^2 ) + \ldots \label{eq:amplitude} \\
\dot A_3 & = & \lambda_2 A_3 + \alpha_2 A_1 A_2 + 
A_3 ( d |A_1|^2 + d |A_2|^2 + e |A_3|^2 ) + \ldots \nonumber,
\end{eqnarray}
where $\lambda_1$ and $\lambda_2$ are the linear growth rates of
the respective modes, and $\alpha_1, \alpha_2, a, b_0, c, d$ and 
$e$ are all real-valued constants.  The quadratic coefficients,
$\alpha_1$ and $\alpha_2$ are non-zero only if the temporal 
resonance condition is met. Note that there is no assumption of
weak forcing/damping and so terms such as $\overline{A_1}$ do
not appear \acs{(}\cite{ANR_SIADS14}\acs{)}.

Now, the mode with wavevector $\bf k_3$ is not at its 
critical point but is
weakly damped, so $\lambda_2<0$ and, close to onset (i.e. $|\lambda_1|$
small enough), it is therefore slaved by
the critical modes. Consequently, 
one can perform a centre manifold reduction on equations
(\ref{eq:amplitude}), resulting in 
$$
A_3 = - \frac{\alpha_2}{\lambda_2} A_1 A_2 + \ldots,
$$
and
\begin{eqnarray}
\dot A_1 & = & \lambda_1 A_1 + 
A_1 ( a |A_1|^2 + b(\theta_{\rm res}) |A_2|^2 ) + \ldots \nonumber \\
\dot A_2 & = & \lambda_1 A_2 + 
A_2 ( b(\theta_{\rm res}) |A_1|^2 + a |A_2|^2 ) + \ldots \label{eq:cmamplitude}, 
\end{eqnarray}
where
\begin{equation}
b(\theta_{\rm res}) = b_0 + b_{\rm res}, 
\qquad b_{\rm res} =  - \frac{\alpha_1 \alpha_2}{\lambda_2}.
\label{eq:bres}
\end{equation}
The presence of the weakly damped mode therefore changes the value of the
cross-coupling coefficient $b(\theta_{\rm res})$
between modes with wavevectors $\bf k_1$ and $\bf k_2$,
offset at an angle $\theta_{\rm res}$. The value of $\theta_{\rm res}$
is determined by the ratio of critical and weakly damped wavenumbers,
but the same idea holds for any three-wave interaction between any $\bf k_1$
and $\bf k_2$ on the critical circle.  This results in a function $b(\theta)$
with either a distinctive peak or dip at $\theta=\theta_{\rm res}$,
depending on the sign of $\alpha_1\alpha_2$ and on the weakness of the 
damping for the weakly damped mode.

Analysing the amplitude equations 
(\ref{eq:cmamplitude}) shows that there are two types of solutions that 
bifurcate from the trivial solution $A_1=A_2=0$, namely stripes 
($A_1 \neq 0, A_2 = 0$ or vice versa) and rectangles ($A_1=A_2$). The 
relative stability 
of rectangles to stripe perturbations is dependent on the relative
size of the self-coupling coefficient $a$ in equation (\ref{eq:cmamplitude}) and 
$b(\theta_{\rm res})$ where, if $b_{\rm res}>0$ ($\alpha_1\alpha_2>0$), 
then the stability
of rectangles is enhanced 
by the three-wave interaction   
and if $b_{\rm res}<0$ ($\alpha_1\alpha_2<0$),
then the stability of
rectangles is suppressed. 

Of course, in the Faraday problem there are not
just two modes and there are many interactions and several weakly
damped circles, but this idea that 
three-wave resonances can promote patterns 
associated with the angle $\theta_{\rm res}$ 
is powerful (\cite{STS_PhysD00}).  
The idea is that the dispersion relation determines which wavenumbers
are critical or close to critical;  allowed three-wave 
resonant interactions then select out particular wavevectors;
the allowed wavevectors are necessarily oriented at  
particular angles as determined by the ratios of available wavenumbers, 
leading to specific values for $\theta_{res}$.  
Superlattice patterns essentially consist of a 
nonlinear superposition of two set\acs{s} of hexagons offset at an
angle to each other.  
There is a 
whole family of different superlattice patterns (\cite{DSS_NL97}),
each corresponding to a different angle, but the particular 
superlattice that will be promoted will be that with an angle
given by $\theta_{res}$.

The superlattice patterns observed by \cite{KPG_PhysD98} for 
$\{6,7\}$ excitation exemplify the idea and motivated the early
theoretical work.
The superlattice patterns that they observe are spatially periodic and 
a spatial Fourier transform of the pattern indicates that they consist 
of essentially
two sets of hexagonal modes offset by an angle of $22^\circ$.
In this case, the particular three-wave resonance of relevance
is of the type illustrated in figure \ref{fig:threewave}(b) and  
is between a harmonic mode that is weakly damped 
of frequency $2 \acs{\tilde \omega}/2$, related to the $2 \acs{\tilde \omega}/2$ tongue, and 
two wavevectors with wavenumbers corresponding
to the $6\acs{\tilde \omega}/2$ tongue (\cite{STS_PhysD00}).  
This was surprising because the superlattice state appears near the 
bicritical
point where both $6 \acs{\tilde \omega}/2$ and 
$7 \acs{\tilde \omega}/2$ onset simultaneously and naively, from 
inspection of the
linear stability diagram (see figure~\ref{fig:tongues}(b)) one would assume
that it was a result of the interaction of $6 \acs{\tilde \omega}/2$ and 
$7\acs{\tilde \omega}/2$ 
modes---the $2\acs{\tilde \omega}/2$ tongue onsets at an amplitude of approximately $50$ so
is not even visible on the scale shown.

This argument not only explains the presence of the $22^\circ$ 
superlattice patterns but also why patterns with angles close to  $30^\circ$ are
seen in $j=\{4,5\}$ forcing (\cite{SG_SIAP07}). Extensions of this basic
idea have been used to
suggest ways to design forcing frequencies to promote particular patterns
in both Faraday waves (\cite{PTS_PRL04}) and
in a model partial differential equation (\cite{RucklidgeSilber_SIADS09}).


%
%
%

The results of \cite{KPG_PhysD98} and the theoretical results
of \cite{STS_PhysD00} showed how 
modes visible in the spatial
Fourier transform of the experimentally observed pattern 
can be linked to critical/weakly damped modes and 
three-wave resonant interactions.
It follows that given an experimentally observed pattern it should
be possible to confirm whether or not it is a result of a particular
interaction by establishing which modes are present in the pattern.
However, identifying precisely which modes are involved can be tricky, as 
we illustrate in the following section. 

\subsection{Practical identification of relevent modes}

\begin{figure}
\begin{center}
\includegraphics[scale=1]{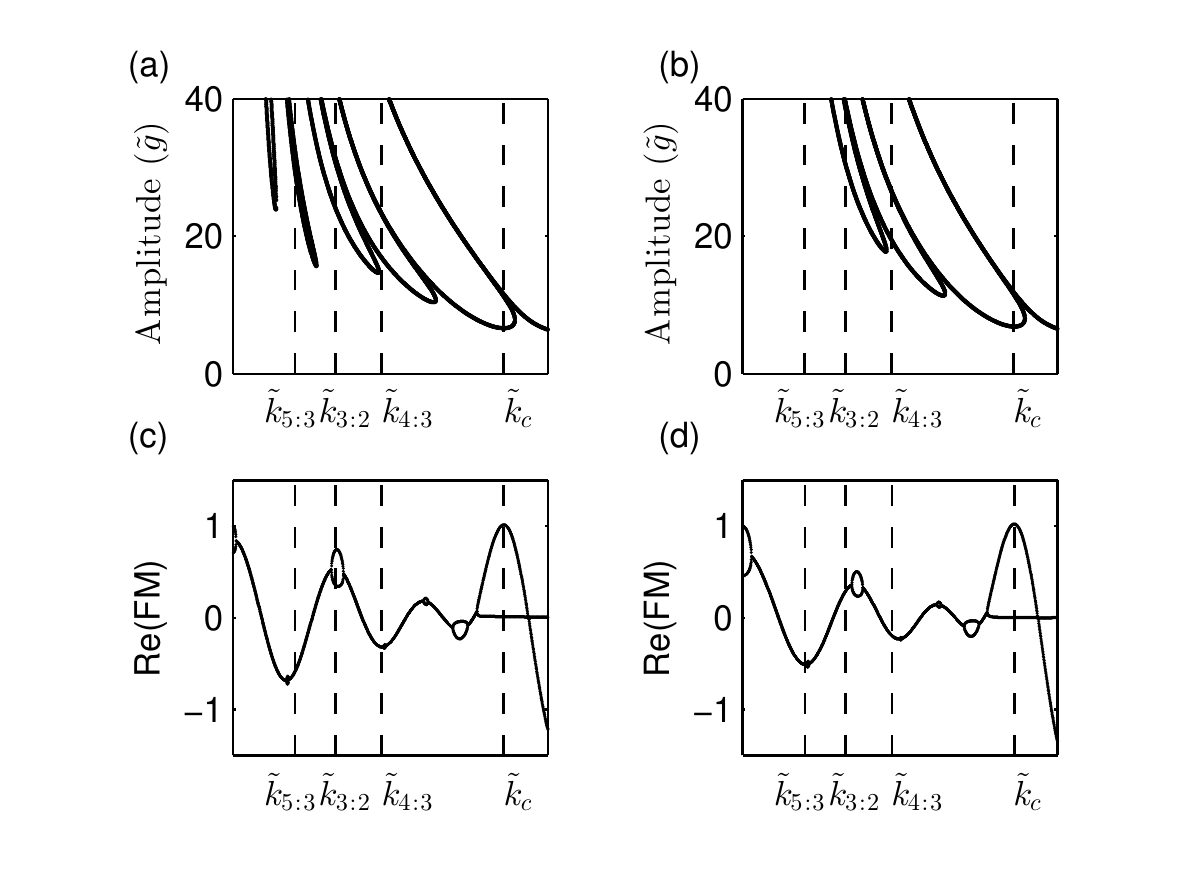}
\end{center}
\caption{Linear stability curves and Floquet multipliers calculated from the 
finite depth Navier--Stokes equations (a) and (c) $\acs{\tilde h}=0.3$ cm 
(b) and (d) $\acs{\tilde h}=0.2$ cm.
Parameter values are taken from \cite{EF_PRE06} and are
$\acs{\tilde \sigma}=20.6$ dyn cm$^{-1}$; 
$\acs{\tilde \nu}=0.18$ cm$^2$ s$^{-1}$; 
$\acs{\tilde \varrho}=0.949$ g m$^{-3}$; 
$\acs{\tilde a_j}, j=\{6,7,2\}$, where $\acs{\tilde a_6} = 3.4, 
\acs{\tilde a_7}=6.4, \acs{\tilde a_2}=0.5$; 
$\phi_j=0^\circ$;
$\acs{\tilde \omega}/2\pi=14$Hz.
}
\label{fig:FMforEF}
\end{figure}

In \cite{EF_PRE06} a detailed comparison of different
patterns, referred to as `grid' states and  
labelled as $3:2$, $4:3$ and $5:3$ respectively is presented. 
The spatial Fourier transform of the patterns enable\acs{s} the identification of the 
wavenumbers that are present in these
patterns: in each case they appear to be dominated by 
modes sitting on two different
circles, one on the primary harmonic instability with wavenumber $\acs{\tilde k_c}$
and a second circle with respective wavenumbers 
$\acs{\tilde k_{3:2}} = 0.38 \acs{\tilde k_c},  
\acs{\tilde k_{4:3}} = 0.55 \acs{\tilde k_c}$ and  
$\acs{\tilde k_{5:3}} =0.23 \acs{\tilde k_c}$. These
ratios are fixed by the fact that the observed grid states 
consist of modes that fit on to a hexagonal lattice.  This
suggests that they are strong candidates for superlattice patterns that
occur as a result of the mechanism discussed in section 4.1.  
\cite{EF_PRE06} computed the linear stability curves, and their 
linear stability diagrams are re-computed in
figure~\ref{fig:FMforEF}(a) and (b).  
\acs{We note that the linear stability curves are similar, but not identical, 
to those given in \cite{EF_PRE06} particularly at small $\acs{\tilde k}$. 
This is because we have made a 
different choice in the sign of the forcing term to that made by
\cite{EF_PRE06}. This is explained further 
in Section \ref{sect:discussion}. 
}
As stated by \cite{EF_PRE06}
and, as can be seen in this figure, 
the $\acs{\tilde k_{3:2}}$ wave number 
lines up with the first harmonic tongue ($2\acs{\tilde \omega}/2$)
and this strongly suggests
that it is the same mechanism that 
produces the familiar $22^\circ$ superlattice patterns, seen by
\cite{KPG_PhysD98}, that
produces the $\acs{\tilde k_{3:2}}$ grid state.
For the two other states the situation is less clear.
\cite{EF_PRE06} state that 
$\acs{\tilde k_{4:3}}$ lines up with the 
the second harmonic tongue ($4\acs{\tilde \omega}/2$), but this is not clear
from figure~\ref{fig:FMforEF} (the position of the 
line drawn in figure $5$ of \cite{EF_PRE06} is not consistent with
the value of $0.55\acs{\tilde k_c}$ given).  \cite{EF_PRE06}
observe that the $\acs{\tilde k_{5:3}}$ mode does not appear to be aligned with any tongue
and that the $5:3$ pattern occurs in a wedge-shaped region of parameter 
space that emerges from the bicritical point, unlike   
the $3:2$ and the $4:3$ state that
occur at a short distance from this point. 

The difficulty here is that the linear stability diagram indicates that at an
amplitude of excitation of approximately $7.5\acs{\tilde g}$ the mode with wavenumber
$\acs{\tilde k_c}$ and frequency $6 \acs{\tilde \omega}/2$ becomes unstable and that there are
five other damped modes with smaller wavenumber.  But the position of the
tongues for these modes only give\acs{s} a rough idea of 
the precise wavenumber and damping associated with each mode at the
pattern onset amplitude of $7.5 \acs{\tilde g}$.  

More accurate information on the damped modes can be obtained by calculating
the most critical Floquet multipliers, as detailed in Appendix \ref{appA}.  
The results for the parameter values
used by \cite{EF_PRE06} are shown in figure~\ref{fig:FMforEF}(c) and (d).  
\acs{We note that with either possible sign for the forcing, the Floquet multipliers
look very similar.}
As identified by \cite{EF_PRE06}, the $\acs{\tilde k_{3:2}}$ mode
is aligned with the harmonic tongue associated with frequency $2\acs{\tilde \omega}/2$ for
$\acs{\tilde h}=0.3$ cm, but not for $\acs{\tilde h}=0.2$ cm and supports their observation that
the $k_{3:2}$ grid state will be seen for $\acs{\tilde h}=0.3$ cm and not for 
$\acs{\tilde h}=0.2$ cm.  
The remaining two wavenumbers $\acs{\tilde k_{5:3}}$ and $\acs{\tilde k_{4:3}}$ are
very close to the subharmonic waves of frequencies $\acs{\tilde \omega}/2$ and
$3 \acs{\tilde \omega}/2$ respectively.  This clarifies the issue that the $\acs{\tilde k_{5:3}}$
mode in \cite{EF_PRE06} did not appear to align with any mode on
the linear stability graphs and  
strongly suggests that the $\acs{\tilde k_{4:3}}$ mode was mis-identified as harmonic.  
However,
it also presents a problem in that temporal
constraints mean that it is not possible for two harmonic wavevectors to
form a three-wave resonance with a single subharmonic wavevector. 
Consequently, $\alpha_1=\alpha_2=0$ in equations (\ref{eq:amplitude}), so 
the discussion in section 4.1 does not hold as it stands 
and there is no special angle $\theta_{\rm res}$.

\cite{EF_PRE06} point out that the $\acs{\tilde k_{5:3}}$ case could be a result of 
a three-wave resonance between modes with frequencies $\acs{\tilde \omega}/2, 
6\acs{\tilde \omega}/2$ 
and $7\acs{\tilde \omega}/2$ and this is strongly supported by our Floquet multiplier calculation
and is a permitted resonance in that it satisfies both spatial and temporal
constraints.  Such a resonance appears to be consistent with the spatial Fourier
transform in figure 3(c) of \cite{EF_PRE06} and is similar in structure to
previous states identified in figure 20(c) of \cite{AF_PRE02} as 2MS states.
This would make it a three-wave resonance of the form shown
in figure \ref{fig:threewave} (d).  This kind of resonance was discussed
briefly by \cite{PS_PRL02} and results in patterns that occur in a wedge
that emerges from the
bicritical point.  This is consistent with the bifurcation sets shown
by Epstein and Fineberg.  The only remaining conundrum for this particular
pattern is that Epstein and Fineberg state that, although all the evidence
points to subharmonic modes being important, no subharmonic component
was found.

\section{Observed patterns near onset: theoretical bifurcation sets 
   compared with experiment}

In spite of the difficulties in identifying modes, nevertheless, in
experiments with dominant forcing frequencies $j=\{4,5\}$ or $j=\{6,7\}$ there is a
coherent picture: in the $\{6,7\}$ case, 
observed superlattice patterns near onset
have an angle of $22^\circ$; for $j=\{4,5\}$, observed patterns 
near onset are quasipatterns with an angle of $30^\circ$. 
Although there are technical difficulties with considering
quasipatterns in the same way as superlattice patterns 
(\cite{RR_PhysD03,RucklidgeSilber_SIADS09,IoossRucklidge_JNonlSci10}), 
nevertheless 
they appear to fit within the same framework with 
resonant interactions with weakly damped
modes associated with $2\acs{\tilde \omega}/2$ contributing to $b_{res}$ 
and explaining the appearance
of the appropriate angle.   The 
different angle seen in the $\{4,5\}$ case as compared with the
$\{6,7\}$ case is because of differences in the wavenumbers of the interacting modes.  In 
both cases,
superlattice/quasipatterns are only seen near the bicritical point
where the two modes driven by the two main components of the forcing
onset simultaneously. 

Here, we aim to carry out a careful comparison of the results
of weakly nonlinear theory and experimental results 
to explore to what extent there is quantitative agreement between experiments
and theory as the bicritical point is approached.  
We present results from the experiments of \cite{DU_PRE06}
because we have
excellent agreement for the linear stability curves
with $\acs{\tilde \nu}=0.21$ cm$^2$ s$^{-1}$, as shown
in figure~\ref{fig:linearstabilitycomparison}(e) and (f), 
and this set of experiments includes the most comprehensive
study of how multiple frequencies interact via the superlattice mechanism
outlined above. Specifically, acknowledging the importance of the $2\acs{\tilde \omega}/2$ 
tongue, \cite{DU_PRE06} 
carry out a series of experiments that systematically explore
the effect of adding a third frequency that promotes this mode.

The weakly nonlinear theory is carried out using the method 
given in \cite{SG_SIAP07} for analysing 
equations (\ref{NSndf}) to \acs{\sout{(2.7)}(\ref{bcbottomp})}. 
For all details of how weakly nonlinear analysis is used to derive
coefficients for the relevant 
amplitude equations we refer the reader to \cite{SG_SIAP07}.  In order to 
examine the relative stability of superlattice patterns a minimal set of twelve
amplitude equations corresponding to twelve vectors on the critical circle
are needed.  However, the stability of the planforms that
bifurcate from the non-patterned state can be found by calculating the coefficients on
three one-dimensional subspaces of this twelve dimensional problem: one describing stripes,  
one for hexagons and one for  rectangles generated from wavevectors 
separated by an angle $\theta$ as detailed in \cite{SS_PhysD98}. 
By varying $\theta$, this formulation then results in
determining the stability of stripes to any perturbation of the same wavenumber; 
the stability of hexagons to (i) stripes arbitrarily
close to any given orientation (ii), rectangles arbitrarily close to any given 
aspect ratio, (iii) to superlattice patterns of any angle; the stability
of rectangles to (i) stripes of any orientation (ii) superlattice patterns
on the same lattice;
the stability of superlattice patterns to hexagons, 
stripes or rectangles that are made up of a subset 
of the superlattice wavevectors.  

Since the weakly nonlinear stability analysis indicates
that for many parameter values there is more than one stable pattern,
results are presented in two ways, first showing
the regions of stability for individual patterns---only those that have some
region of stability are shown,  and secondly showing the
bifurcation set for the most stable state, as computed from the weakly
nonlinear coefficients using the Lyapunov 
functional given in \cite{SG_SIAP07}.  The most stable states are  
superimposed on the experimental results of \cite{DU_PRE06}.  
In the case of superlattice patterns, the resonant interactions mean
that it is always the particular pattern associated with
$\theta_{res}$ that has the largest region of stability and it 
is perturbations associated with this angle that first destabilise 
hexagons. Consequently where superlattice patterns are shown it is
always the superlattice pattern associated with $\theta_{res}$ that is relevant.

\begin{figure}
\begin{center}
\includegraphics[scale=1.0]{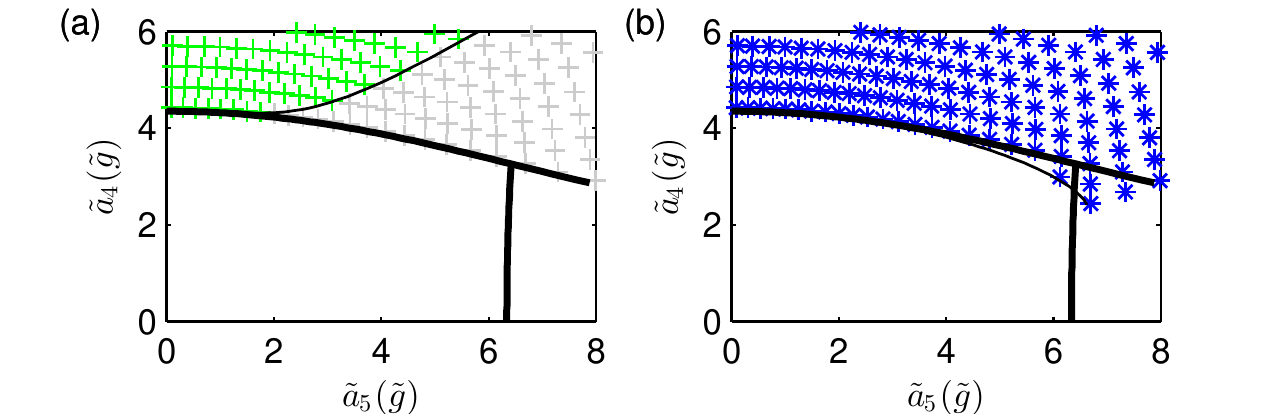}

\includegraphics[scale=1.0]{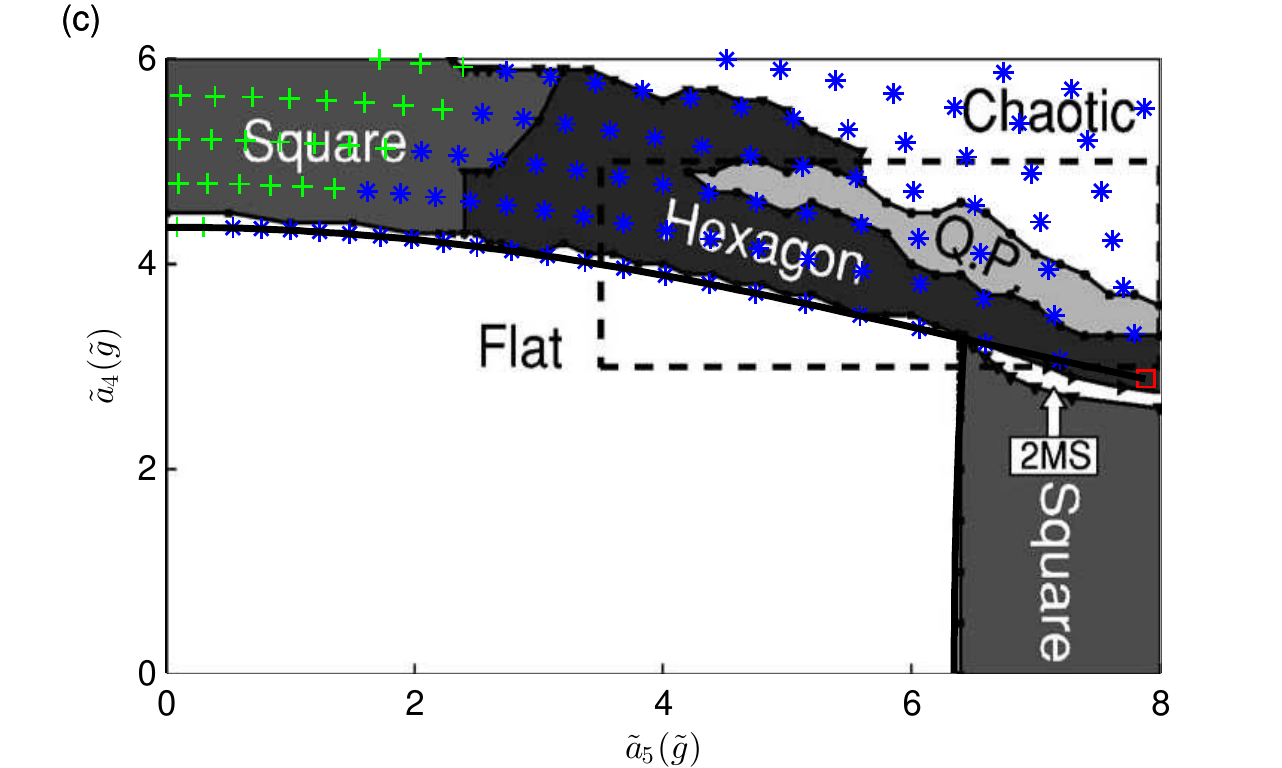}

\end{center}
\caption{Comparison with figure 1 from \cite{DU_PRE06}. 
$\acs{\tilde a_j}=\{4,5\}, (\phi_4, \phi_5) = (0,16^\circ), 
\acs{\tilde \omega}/2\pi = 20$ Hz.  The
fluid parameters are as in the caption to figure 3(e),(f) with
$\acs{\tilde \nu}=0.21$ cm$^{2} s^{-1}$.
(a)-(b) Stability regions for different planforms.  
(a) Stable squares in green; unstable squares in grey; (b) stable
hexagons in blue; unstable hexagons in grey. \acs{Note that 
for $\acs{\tilde a_5}>0$, there are quadratic terms in the amplitude equations which
mean that hexagons bifurcate transcritically. The unstable solutions
that are produced in this transcritical bifurcation are stabilised in a
saddle-node bifurcation, resulting in a small region of stable, 
subcritical hexagons. This region of stable subcritical hexagons is only
visible on the scale of the figure close to the bicritical point.}
(c) `Most stable' state using a Lyapunov energy argument: $+$ squares;
$*$ hexagons; $\Box$ quasipatterns.  The theoretical results
are overlaid on figure 1 from \cite{DU_PRE06}.
Adapted with permission from \cite{DU_PRE06}.  Copyrighted by
the American Physical Society.
}
\label{fig:DU_fig1}
\end{figure}

Ding and Umbanhowar consider the two sets of frequency components most
widely used by other experimental groups, namely $j=\{4,5\}$ and $j=\{6,7\}$ and
then systematically investigate the inclusion of a third mode, so 
considering the combinations $j=\{4,5,2\}$ and $j=\{6,7,2\}$.  We have carried
out the weakly nonlinear stability analysis for 
each of their parameter studies and the results are presented in
figures~\acs{\sout{7}\ref{fig:DU_fig1}}-\ref{fig:DU_fig6b} where the principle parameters considered
and the corresponding figures in \cite{DU_PRE06} are summarised in table~\ref{tab:figsummary}.

\begin{table}
\begin{center}
\begin{tabular}{ccccc}
\def~{\hphantom{0}}
Figure             & $j$         & $\acs{\tilde a_j}, \phi_j$   & Parameters      & Figure from  \\ 
                   &             &                 & varied          & \cite{DU_PRE06} \\[3pt] 
\ref{fig:DU_fig1}  &  $\{4,5\}$  & $(\phi_4,\phi_5)=(0,16^\circ)$
                                                   & $\acs{\tilde a_4}$ and $\acs{\tilde a_5}$ & 1 \\
\ref{fig:DU_fig3b} & $\{4,5,2\}$ & $\acs{\tilde a_2}=0.8\acs{\tilde g}$, 
             $(\phi_4,\phi_5,\phi_2)=(0^\circ,16^\circ, 32^\circ)$ 
                                                   & $a_4$ and $a_5$ & 3 (b) \\
\ref{fig:DU_fig4a} & $\{4,5,2\}$ & $\acs{\tilde a_5}=5\acs{\tilde g}$, 
                         $(\phi_4,\phi_5,\phi_2)=(0^\circ,5^\circ, 32^\circ)$ 
                                                   & $\acs{\tilde a_4}$ and $\acs{\tilde a_2}$ & 4 (a) \\
\ref{fig:DU_fig4b} & $\{4,5,2\}$ & $(\acs{\tilde a_4,\tilde a_5,\tilde a_2})=(3.8,5.8,0.8)g, \phi_4=0^\circ$ 
                                                   & $\phi_5$ and $\phi_2$  
						                     & 4(b) \\
\ref{fig:DU_fig2}  & $\{6,7$ & $(\phi_6,\phi_7)=(0^\circ, 40^\circ)$ 
                                                   & $\acs{\tilde a_6}$ and $\acs{\tilde a_7}$ & 2 \\
\ref{fig:DU_fig6a} & $\{6,7,2\}$ & $\acs{\tilde a_7}=7\acs{\tilde g}, (\phi_6,\phi_7,\phi_2)=(0^\circ, 40^\circ,80^\circ)$ 
                                                   & $\acs{\tilde a_6}$ and $\acs{\tilde a_2}$ & 6(a) \\
\ref{fig:DU_fig6b} & $\{6,7,2\}$ & $(\acs{\tilde a_6,\tilde a_7,\tilde a_2})=(5.2,7.8,0.6)
 \acs{\tilde g}$, $\phi_6=0^\circ$ 
                                                   & $\phi_7$ and $\phi_2$  & 6(b)\\
\end{tabular}
\caption{Summary information for figures comparing weakly nonlinear analysis with
the experiments of \cite{DU_PRE06}.}\label{tab:figsummary}
\end{center}
\end{table}

In each case the weakly nonlinear results are superimposed on the experimental
results.  The parameters have been chosen to match those quoted in 
the paper of \cite{DU_PRE06} with the exception of the viscosity where, as
seen in figure~\ref{fig:linearstabilitycomparison}, the linear stability curves
fitted best with a viscosity of 
$\acs{\tilde \nu}=0.21$ cm$^2$s$^{-1}$ rather than 
$\acs{\tilde \nu}=0.20$ cm$^2$s$^{-1}$ quoted in their paper (or the value of 
$\acs{\tilde \nu}=0.204$ cm$^2$s$^{-1}$ quoted in \cite{D_THESIS06}).

\subsection{Results for \{4,5\} and \{4,5,2\}}
In figure~\ref{fig:DU_fig1} we see that
the weakly nonlinear theory predicts bistability between squares and
hexagons for low values
of $\acs{\tilde a_5}$, with squares losing stability as the bicritical point is approached.  
There is good agreement between the point at which squares become unstable
(approximately $\acs{\tilde a_5}=2.2\acs{\tilde g}$, figure \ref{fig:DU_fig1}(a)), 
and the transition between squares and hexagons in
the experiment.  However, before squares become unstable, the theory
suggests that there is a region of bistability between hexagons and squares and
that in this region hexagons and not squares, as seen in the experiment,
are the most stable state.  Of course,
in this region, the particular pattern observed experimentally will depend to some
extent on the way in which the experiments were carried out.  
For example, if the experimental procedure was to fix $\acs{\tilde a_4}$ 
and to increase $\acs{\tilde a_5}$
from zero in small steps, one would expect to see squares until the limit of their
stability.  Regions of quasipatterns
are also predicted by the weakly nonlinear theory (almost off the scale of the figure
\ref{fig:DU_fig1}) and observed in the experiment
but since they do not occur close to onset it 
is perhaps not surprising that the theory and experiment do not agree.

\begin{figure}
\begin{center}
\includegraphics[scale=1.0]{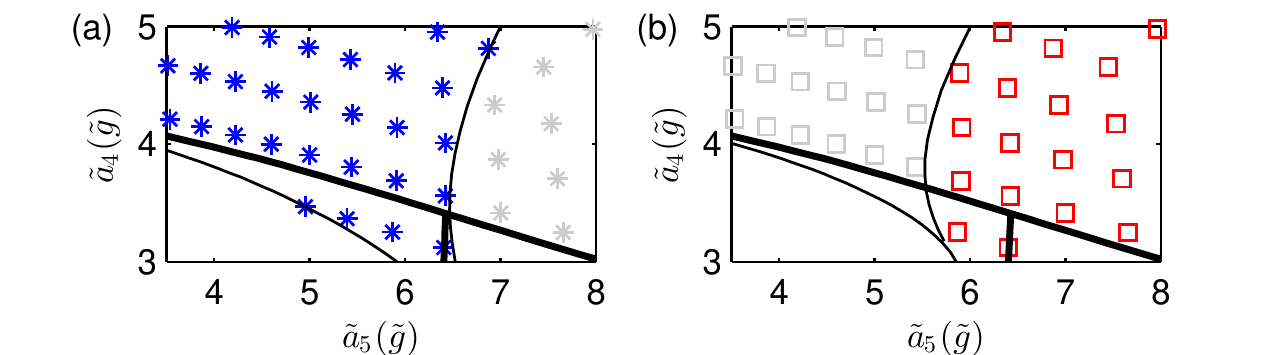}

\includegraphics[scale=1.0]{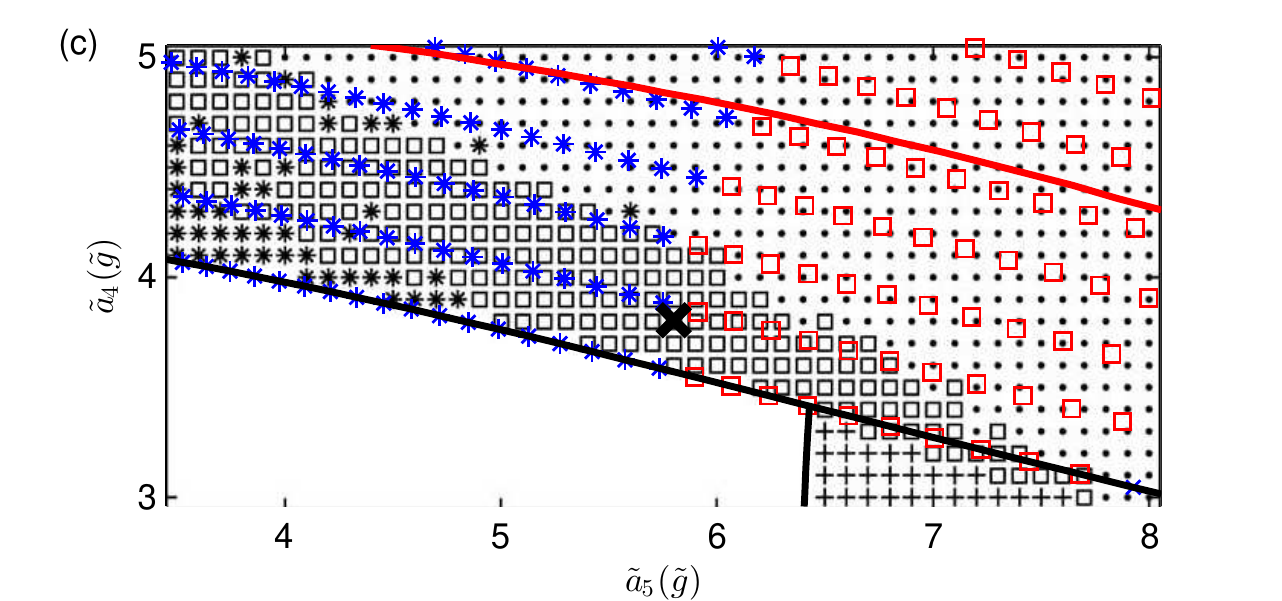}
\end{center}
\caption{Comparison with \cite{DU_PRE06} Figure 3(b), 
$j=\{4,5,2\}, (\phi_4, \phi_5, \phi_2) =( 0, 16^\circ,32^\circ)$. 
(a)-(b) Stability regions for different planforms.  
(a) Stable hexagons in blue, unstable hexagons in grey; (b) stable
quasipatterns in red, unstable quasipatterns in grey. 
\acs{Both
quasipatterns and hexagons bifurcate transcritically and may be
stabilised in a saddle-node bifurcation.  This can result in 
stable subcritical quasipatterns/hexagons.}   
(c) `Most stable' planform using a Lyapunov energy argument 
superimposed on figure 3(b)
from \cite{DU_PRE06}. 
Adapted with permission from \cite{DU_PRE06}.  Copyrighted by
the American Physical Society.
$*$ hexagons; $\Box$ quasipatterns, $\bullet$ disordered states,
where the coloured symbols are the theoretical
results and the black symbols the experimental results.  
The red line is the linear instability curve
for the harmonic $2\acs{\tilde \omega}/2$ tongue. The black cross indicates the point at which
the experiment shown in figure~\ref{fig:DU_fig4b}(c) was carried out.
}
\label{fig:DU_fig3b}
\end{figure}

In figure~\ref{fig:DU_fig3b}, we see that 
by adding in a forcing component that excites the $2 \acs{\tilde \omega}/2$ mode 
directly \cite{DU_PRE06}
showed that they could increase the region of quasipatterns, to the extent
that they become the first pattern observed after the unpatterned state becomes unstable.
Our theoretical results for the same parameter values also show an enhanced region
for the superlattice patterns---but not as enhanced as for the experiment.  Two
possible reasons for this are (i) the nearby presence of the bicritical point
where both the $2\acs{\tilde \omega}/2$ and the 
$4\acs{\tilde \omega}/2$ tongue onset simultaneously which
means that our weakly nonlinear calculations would need to be extended;
(ii)
the sensitivity of the results to phase.  We discuss both of these in more detail below.

\begin{figure}
\begin{center}
\includegraphics[scale=1.0]{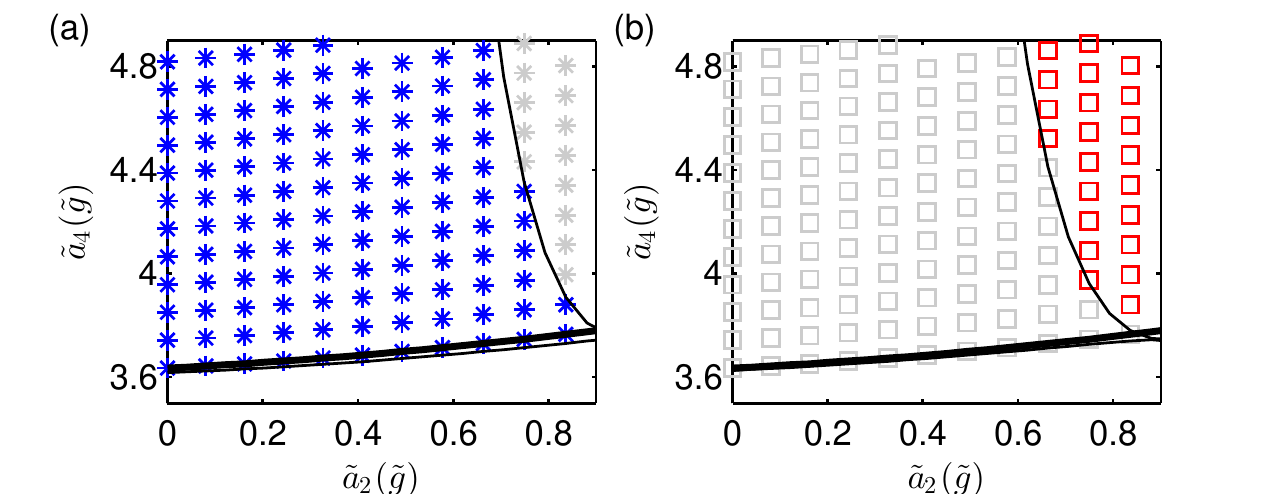}

\includegraphics[scale=1.0]{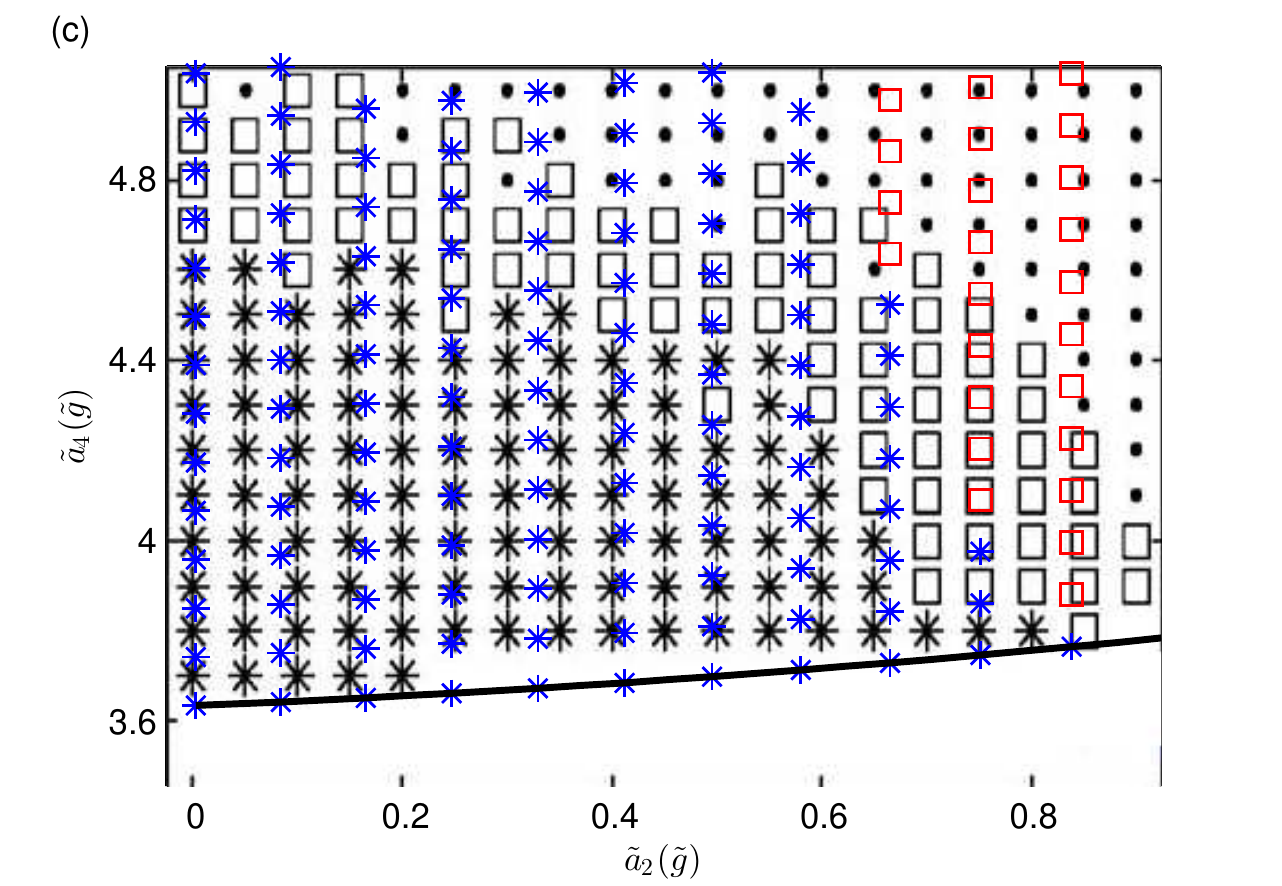}

\end{center}
\caption{Comparison with \cite{DU_PRE06} figure 4(a). 
$j=\{4,5,2\}, \acs{\tilde a_5}=5\acs{\tilde g}, 
(\phi_4, \phi_5, \phi_2) =  (0, 16^\circ, 32^\circ).$
(a) and
(b) stability regions for hexagons and quasipatterns respectively.
Coloured symbols stable; grey unstable.  Hexagons and quasipatterns
both bifurcate from the unpatterned state at a transcritical
bifurcation, where the unstable bifurcating branch changes direction
at a saddle-node bifurcation.  This leads to regions of stable hexagons
and quasipatterns that are subcritical.
(c) `Most stable' planform: 
$*$ hexagons; $\Box$ quasipatterns, overlaid on figure 4(a)
from \cite{DU_PRE06}. 
Adapted with permission from \cite{DU_PRE06}.  Copyrighted by
the American Physical Society.
The coloured symbols are the theoretical
results and the black symbols the experimental results.  
The black dots indicate disordered states.
Note that the linear stability boundary for the 
$2\acs{\tilde \omega}/2$ mode is just off this diagram, and is 
approximately parallel to the $a_4$ axis
through a value of $a_2 \approx 1$.  The bicritical point
where the $2 \acs{\tilde \omega}/2$ and the $4\acs{\tilde \omega}/2$ modes onset 
simultaneously occurs
when $(\acs{\tilde a_2, \tilde a_4})=(1.05,3.80)\acs{\tilde g}$.
}
\label{fig:DU_fig4a}
\end{figure}

Figure~\ref{fig:DU_fig4a} shows
excellent agreement between the theoretical results and the experimental results
for the pattern at onset for increasing $\acs{\tilde a_2}$, with agreement diminishing 
with distance from onset.

\begin{figure}
\begin{center}
\includegraphics[scale=1.0]{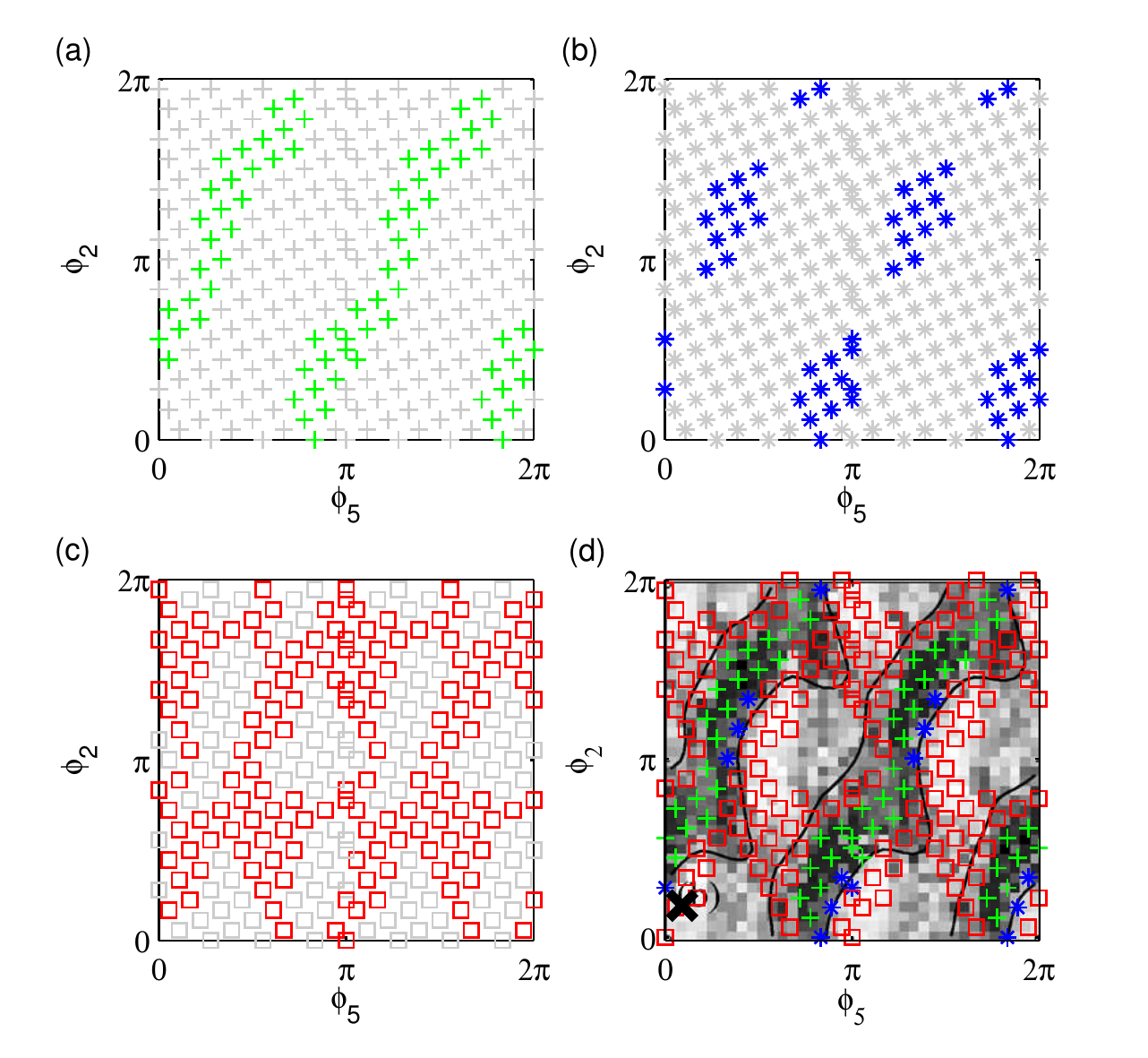}

\end{center}
\caption{Comparison with \cite{DU_PRE06} figure 4(b). 
Phase diagram for $j=\{4,5,2\}$ with $(\acs{\tilde a_4,\tilde a_5,\tilde a_2})=(3.8,5.8,0.8)\acs{\tilde g}$.
(The position of this point for $(\phi_5, \phi_2) = ( 16^\circ, 32^\circ)$ is
shown in figure~\ref{fig:DU_fig3b}(c).)
(a)-(c) 
stability regions for squares, hexagons and quasipatterns respectively.
Coloured symbols stable; grey unstable.  (d) `Most stable' planforms: 
$+$ squares; $*$ hexagons; $\Box$ quasipatterns.  The
results are overlaid on
figure 4(b) of \cite{DU_PRE06}. 
Adapted with permission from \cite{DU_PRE06}.  Copyrighted by
the American Physical Society.
The cross indicates the point which
corresponds to the cross in figure~\ref{fig:DU_fig3b}.
 }
\label{fig:DU_fig4b}
\end{figure}

There is surprisingly good agreement for the variation with phase
shown in figure~\ref{fig:DU_fig4b}.  In the experimental
results, black indicates low correlation at an angle of $30^\circ$.  This
lines up well with areas where the theory indicates competition between
squares and hexagons.  White areas, where there is high correlation at
$30^\circ$, line up well with stable quasipatterns.  In the regions
where there are no coloured symbols, the weakly nonlinear analysis indicates 
no stable states.  This region is mostly mid-grey in the experiment, indicating
some correlation at $30^\circ$. 

The agreement in figure~\ref{fig:DU_fig4b} is 
surprising because
the results correspond to a point with $\acs{\tilde a_4}=3.8\acs{\tilde g}, \acs{\tilde a_5}=5.9\acs{\tilde g}$ 
and $\acs{\tilde a_2}=0.8\acs{\tilde g}$,
which according to the theory is close to a transition from quasipatterns
to hexagons in a region where both hexagons and quasipatterns are stable,
whereas in the experiment it is firmly in the quasipattern region.
(In order to help cross-reference between the two figures, a black
cross has been marked on both figure~\ref{fig:DU_fig3b}(c) and
figure \ref{fig:DU_fig4b}.) 
Figure~\ref{fig:DU_fig4b} highlights the sensitivity of the results 
to phase and in fact,
with our weakly nonlinear analysis we find that the size of the
region of quasipatterns
changes significantly depending on the phase.  This sensitivity 
to the precise value of the phase could be part of the difficulty in 
obtaining agreement for the onset of quasipatterns in 
figure~\ref{fig:DU_fig3b}.  

\acs{The results are, of course, also sensitive to the viscosity.  
Away from the bicritical point,
changing the viscosity tends to translate the nonlinear bifurcation lines as one might expect:
upwards if the viscosity is increased, downwards if the viscosity is decreased, 
in line with the changes to the transition from unpatterned to patterned state seen in Figure 3.
Close to the bi-critical point the effects are more
pronounced with small changes in viscosity sometimes leading to large changes in the 
regions in which particular patterns dominate.}

\subsection{Results for \{6,7\} and \{6,7,2\}}
\begin{figure}
\begin{center}
\includegraphics[scale=1.0]{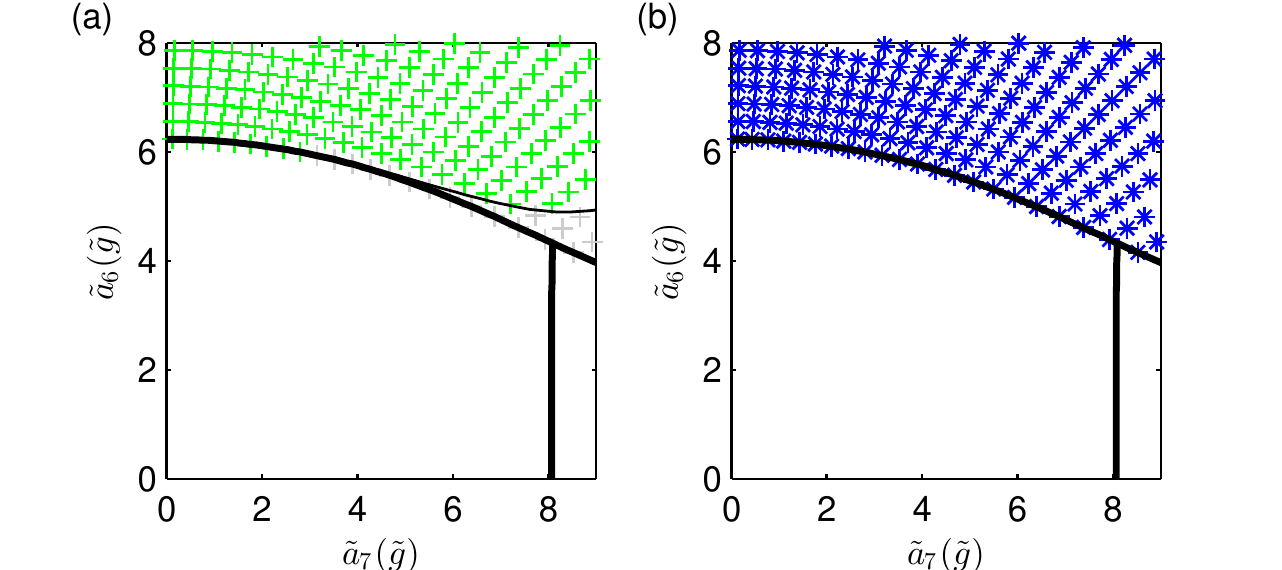}

\includegraphics[scale=1.0]{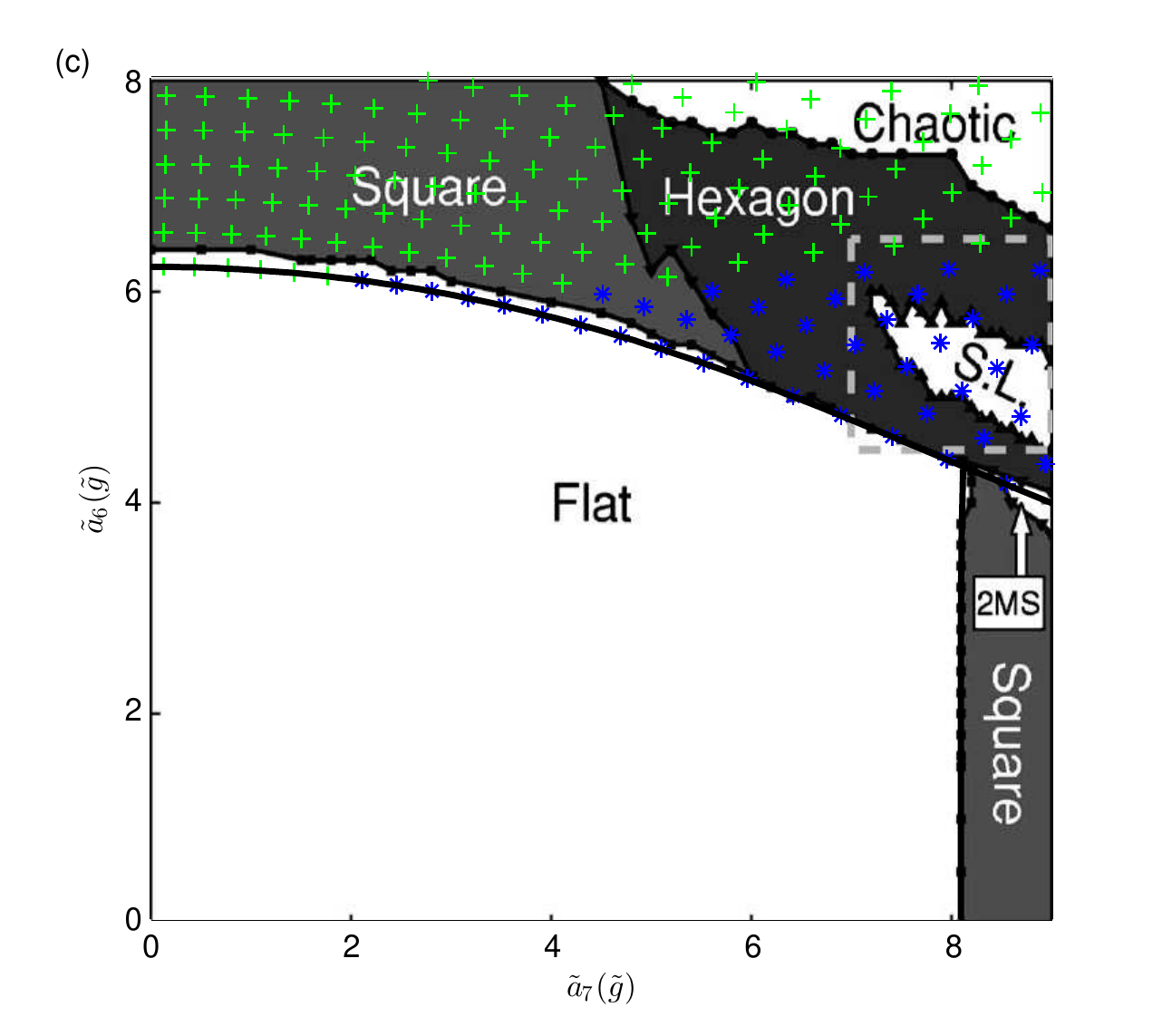}

\end{center}
\caption{Comparison with \cite{DU_PRE06}, figure 2. 
$j=\{6,7\}, (\phi_6, \phi_7)=(0,40^\circ), \acs{\tilde \omega}/2\pi = 16.5$ Hz.
(a)-(b) 
stability regions for rectangles and hexagons.
Coloured symbols stable; grey unstable.  (c) `Most stable' planform: 
$+$ rectangles; $*$ hexagons, overlaid on figure 2 of \cite{DU_PRE06}.
Adapted with permission from \cite{DU_PRE06}.  Copyrighted by
the American Physical Society.
 }
\label{fig:DU_fig2}
\end{figure}

In figure~\ref{fig:DU_fig2}
the weakly nonlinear theory predicts bistability between squares and
hexagons for low values
of $\acs{\tilde a_7}$, with squares losing stability as the bicritical point is approached.  
As for figure~\ref{fig:DU_fig1}, although there is good agreement between the 
point at which squares become unstable
(approximately $\acs{\tilde a_7}=4.5\acs{\tilde g}$), and the transition between squares and hexagons in
the experiment,  this agrees less well with the results of the energy argument.

\begin{figure}
\begin{center}
\includegraphics[scale=1.0]{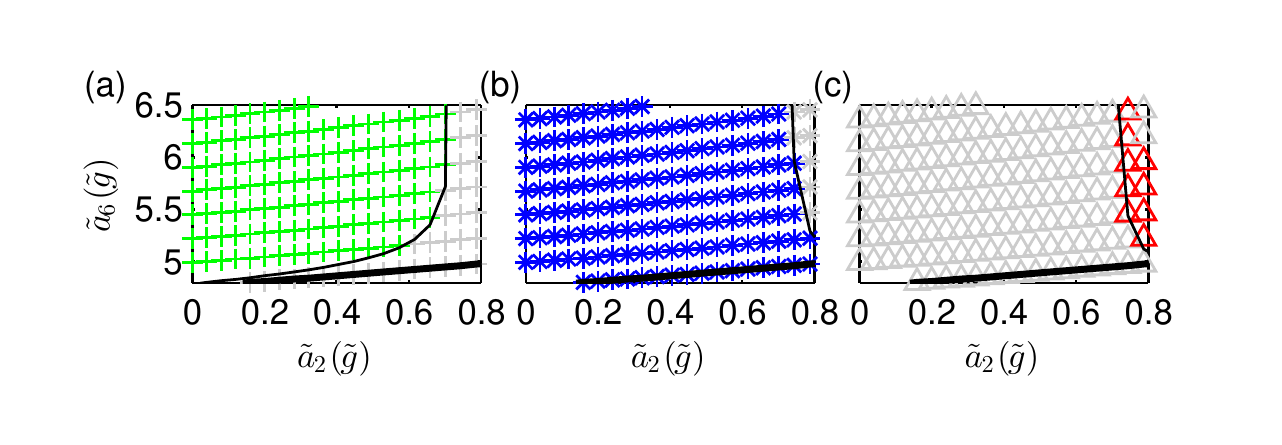}

\includegraphics[scale=1.0]{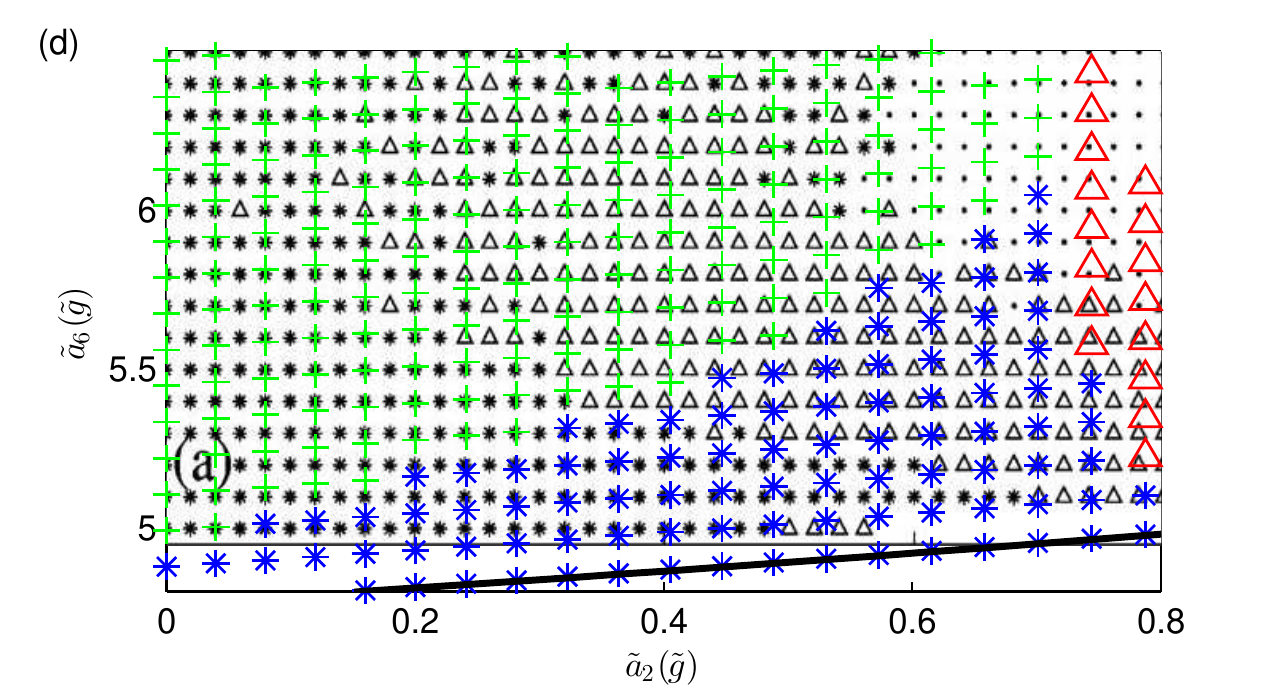}

\end{center}
\caption{Comparison with \cite{DU_PRE06}, figure 6(a). 
$j=\{6,7,2\}, \acs{\tilde a_7}=7\acs{\tilde g}, 
(\phi_6, \phi_7, \phi_2)= (0, 40^\circ, 80^\circ), 
\acs{\tilde \omega}/2 \pi=16.5$ Hz.
(a)-(c) 
stability regions for rectangles, hexagons and superlattice patterns respectively.
Coloured symbols stable; grey unstable.  
(d) `Most stable' planforms superimposed on figure
6(a) from \cite{DU_PRE06}. 
Adapted with permission from \cite{DU_PRE06}.  Copyrighted by
the American Physical Society.
The same symbol styles are used for both experiments
and theory ($\bigtriangleup$ superlattice; $*$ hexagons; 
$+$ rectangles; $\bullet$ disordered states).  Coloured symbols represent theoretical
results and black symbols are the experimental results.
}
\label{fig:DU_fig6a}
\end{figure}

In figure~\ref{fig:DU_fig6a}
both theory and experiment agree that for low and moderate values of $\acs{\tilde a_2}$ 
hexagons are the preferred pattern.  As $\acs{\tilde a_2}$ increases, hexagons
are replaced by superlattice patterns, although the transition observed
experimentally occurs slightly earlier than in the weakly nonlinear
calculations.  Away from onset, the agreement is qualitative rather
than quantitative: in the
experiments superlattice patterns are observed for a much larger region than
predicted by the weakly nonlinear theory and hexagons are observed in regions
where the theory predicts bistability of rectangles and hexagons, with
rectangles being the most stable state.  For large values of $\acs{\tilde a_2}$
and $\acs{\tilde a_6}$, the experiments see disordered states. The weakly nonlinear
theory cannot predict such states, but note that none of the planforms
considered are found to be stable for large $\acs{\tilde a_2}$ and $\acs{\tilde a_6}$ (there are
no symbols from the theoretical calculations in the top right of
figure~\ref{fig:DU_fig6a}).

\begin{figure}
\begin{center}
\includegraphics[scale=1.0]{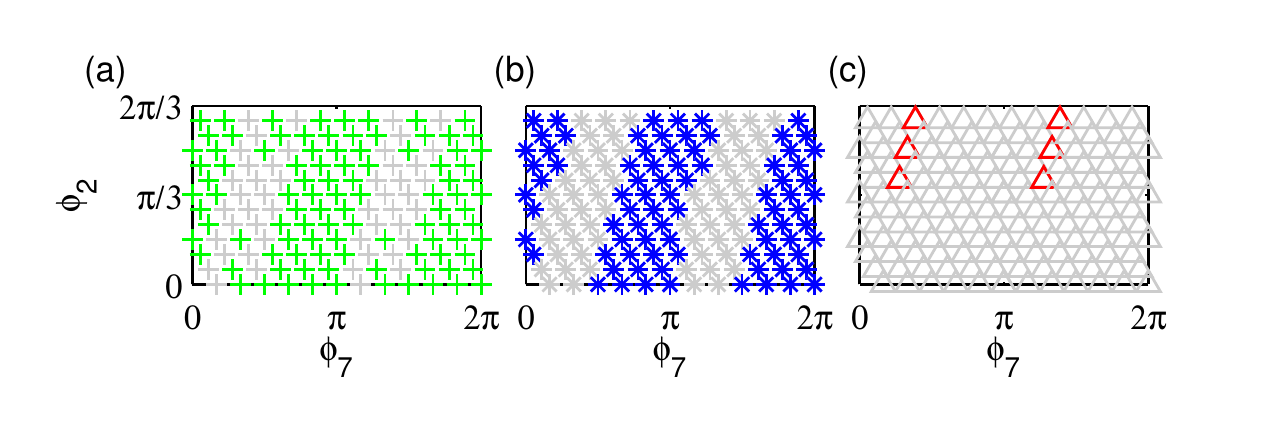}

\includegraphics[scale=1.0]{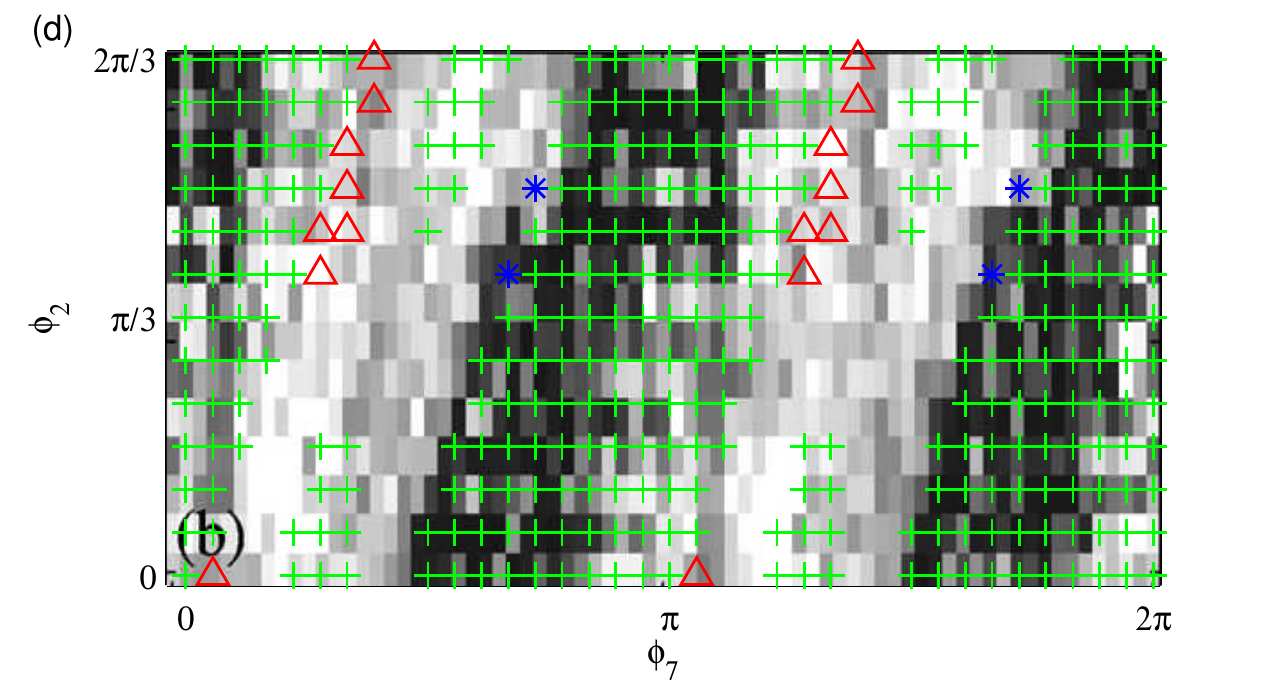}
\end{center}
\caption{Comparison with \cite{DU_PRE06}, figure 6(b).  
Phase diagram for $j=\{6,7,2\}, (\acs{\tilde a_6,\tilde a_7,\tilde a_2})=(5.2,7.8,0.6)\acs{\tilde g}, 
\acs{\tilde \omega}/2\pi=16.5$ Hz. 
(a)-(c) 
stability regions for rectangles, hexagons and superlattice\acs{ patterns} respectively.
Coloured symbols stable; grey unstable.  
(d) `Most stable' planform, superimposed on figure 6(b), \cite{DU_PRE06}. 
$\bigtriangleup$ superlattice; $*$ hexagons; $+$ rectangles.  
Coloured symbols represent theoretical
\acs{results and black symbols are the experimental results.}
Note that in (d) hexagons are rarely the `most stable' planform: whenever
hexagons are stable, rectangles usually are too.  Although the
Lyapunov energy for hexagons is often similar to that for rectangles, 
rectangles nearly always have the lowest value.
}
\label{fig:DU_fig6b}
\end{figure}

The results shown in figure~\ref{fig:DU_fig6b}
again show that the theoretical results have the same diagonal dependency
as seen in the experiments, a consequence of a phase invariant as discussed
further below.
For the experimental results, white indicates a high correlation
with the angle of $22^\circ$ and this does seem to align with
where superlattice patterns occur, although for much of the lighter regions
the weakly nonlinear analysis predicts no stable pattern.  The black regions
correspond to a low correlation at $22^\circ$ and
\cite{DU_PRE06} state that this region contains `disordered' patterns: 
our theoretical
results suggest that both rectangles and hexagons are stable for much of this
region each having a similar Lyapunov energy.  Consequently one might expect
competition between these two states, resulting in the observed disorder.

\subsection{Common themes}
From our comparison, a number of themes emerge:
\begin{enumerate}
\item  There is excellent
agreement with theory at onset in some cases (figure~\ref{fig:DU_fig4a}). 
\item In other cases, figure~\ref{fig:DU_fig1} and \ref{fig:DU_fig2},  
bistability of patterns makes it difficult to compare theory and
experiment.  The experimental results for the 
pattern at onset are consistent with the theory, but sometimes the pattern
at onset is not that predicted by considering the Lyapunov
argument. 
This is one difficulty in comparing with experiment {\em post festum},
 since in regions of bistability, which pattern is observed
can depend on the way that the experiments are performed.
\item When the additional forcing component ($\acs{\tilde a_2} \neq 0$) 
is included that promotes the mode with
frequency $2\acs{\tilde \omega}/2$, then in both theory and experiments 
patterns associated with the resonant angle are promoted and consequently
appear for a larger region of parameter space.
\item In the experimental results,
the additional forcing causes regions
where both hexagonal states and superlattice patterns occur close to onset.  
This is also true in the weakly nonlinear analysis, but to a lesser extent,
see figure~\ref{fig:DU_fig3b} and \ref{fig:DU_fig6a}.
There are two possible causes here: firstly, the 
theoretical results are sensitive to the phase, so changing the phase can
expand the region of stable superlattice patterns.  
Secondly, the presence of the weakly damped 
$2 \acs{\tilde \omega}/2$ mode means that the results are in a region that is close to
a bicritical point where both $2 \acs{\tilde \omega}/2$ and $4\acs{\tilde \omega}/2$ modes
onset simultaneously.  In the neighbourhood of this bicritical point, one
would expect the regions of stability as predicted from
a codimension one analysis, at best, be perturbed, and, at worst, be inaccurate.
We will return to this point in the discussion and note 
for now the
red line in figure~\ref{fig:DU_fig3b} that marks the onset of this mode.
\item Quantitative agreement in most cases only occurs close to onset.  
\end{enumerate}

In addition, in all cases, close to the subharmonic instability boundary 
the theory
predicts squares (not shown in the figures), as seen in the experiments.

\section{Discussion} \label{sect:discussion}
The Faraday problem is an important pattern-forming system, yet it is particularly
challenging to tie theory to experiment.  The experiments are difficult to perform;
the parameter regime of interest (large box, moderate viscosity) along
with the technical difficulties of solving 
the free boundary Navier--Stokes equations
make numerical solution of the problem hard, to the extent that 
although there
has been some progress (see \cite{PJT_JFM09,PJT_PRL12}) it is only very recently
that any fully three-dimensional 
calculations 
of superlattice patterns in the Faraday problem have been reported
\cite{KPTSCJ_15}. 
The fact that the instabilities 
result in an entire circle
of unstable wavenumbers presents considerable theoretical difficulties 
(see \cite{Melbourne_TransAmMatSoc99})
and
has meant that theory is, by necessity, restricted to a
finite number of modes---the finite number normally chosen as the minimal set which will
allow for the existence of the experimental pattern of most interest.  
Furthermore, by its nature, weakly nonlinear theory cannot be expected to 
capture behaviour that is inherently strongly nonlinear: so a weakly 
nonlinear analysis of the transition from non-patterned to patterned state
can only be expected to agree with experiment close to onset.

Nevertheless, our detailed comparison with \cite{DU_PRE06} shows very good agreement
between experiment and weakly nonlinear results at
onset, with the caveat that the theory often predicts bistability of patterns. 
Where there is bistability, the observed pattern will be dependent on the
path taken through parameter space and might consist of competition
between the different possible stable states.
Consequently, in some cases it is only possible to say that the 
weakly nonlinear theory is consistent with experiments.  

The results suggest that the qualitative idea that the three-wave resonances 
determine the angles that appear in the patterns does indeed explain 
many of the patterns observed close to onset. 

\begin{figure}
\begin{center}
\includegraphics[scale=1.0]{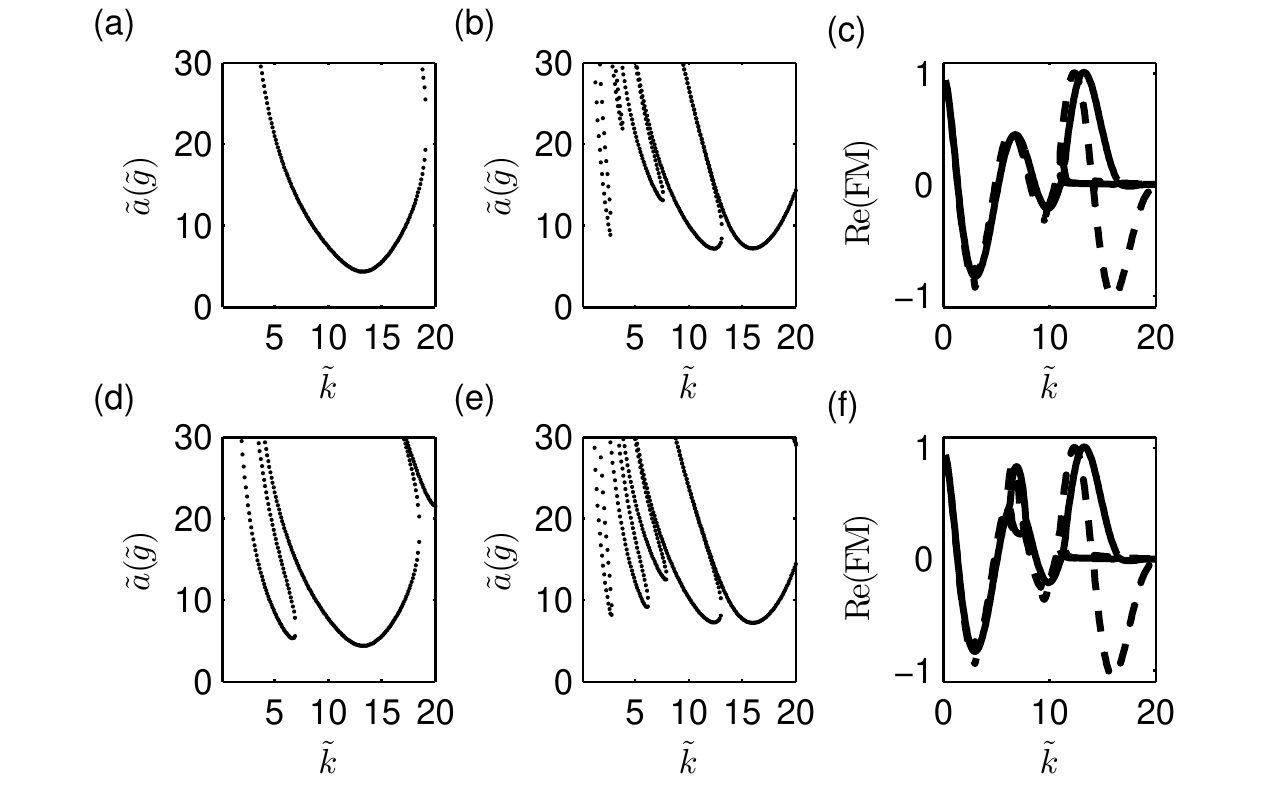}
\end{center}
\caption{Linear stability curves and Floquet multipliers for
$j=\{4,5\}$ and $j=\{4,5,2\}$ excitation using parameter
values as in \cite{DU_PRE06}.  
(a) and (b)  Linear stability curves
for $\acs{\tilde f}(\acs{\tilde t}) = \acs{\tilde a} ( \cos \chi \cos 4 
\acs{\tilde \omega} \acs{\tilde t} + \sin \chi \cos 5 \acs{\tilde \omega} \acs{\tilde t})$, 
where (a) has $\chi=0$ and (b) has $\chi=63^\circ$. (c) the corresponding
Floquet multipliers where the solid line is for $\chi=0$ and the dashed
line is for $\chi=63^\circ$.
\acs{
The Floquet
multipliers are calculated for $\acs{\tilde a}$ fixed at the value for the minimum
of the $4\omega/2$ tongue. So, for $\chi=0, \acs{\tilde a}=4.35\acs{\tilde g}$, 
for $\chi=63^\circ, \acs{\tilde a}=7.19\acs{\tilde g}$
}
(d) and (e)  Linear stability curves
for $\acs{\tilde f}(\acs{\tilde t}) = 
\acs{\tilde a} ( \cos \chi \acs{\cos 4} \acs{\tilde \omega} \acs{\tilde t}
  + \sin \chi \cos 5 \acs{\tilde \omega} \acs{\tilde t} 
  + \acs{\hat a_2} \cos 2 \acs{\tilde \omega} \acs{\tilde t} \acs{)}$ 
where (a) has $\chi=0, a_2 = a \tilde a_2 =0.8g$ 
and (b) has $\chi=62^\circ, \acs{\tilde a_2} = \acs{\tilde a} \acs{\hat a_2} =0.8\acs{\tilde g}$.
(c) 
The corresponding
Floquet multipliers where the solid line is for case (d) where $\acs{\tilde a_5}=0$ 
and the dashed
line is for (e) where $\acs{\tilde a_5} \neq 0$. 
\acs{
The Floquet
multipliers are calculated for $\acs{\tilde a}$ fixed at the value for the minimum
of the $4\acs{\tilde \omega}/2$ tongue. So, for $\chi=0, \acs{\tilde a}=4.40\acs{\tilde g}$, 
for $\chi=63^\circ, \acs{\tilde a}=7.27\acs{\tilde g}.$
}
}
\label{fig:lscandFM}
\end{figure}

The results also highlight the particular aspects of the Faraday
problem that lead to an
amplification of $b_{res}$.  Equation~(\ref{eq:bres}), 
$$
b(\theta_{\rm res}) = b_0 + b_{\rm res}, 
\qquad b_{\rm res} =  - \frac{\alpha_1 \alpha_2}{\lambda_2},
$$
suggests that there are two 
main contributions, one from the quadratic coupling coefficients 
($\alpha_1$ and $\alpha_2$) and one from the
linear damping of the weakly damped mode ($\lambda_2$).  
Promotion of patterns with angle $\theta_{res}$ requires the quadratic
coefficients to be sufficiently large and/or the damping to be sufficiently
small.  As illustrated in figure~\ref{fig:lscandFM}(a) and (b) 
for $j=\{4,5\}$ and in (d) and (e) for $j=\{4,5,2\}$
although increasing $a_5$ and heading towards the bicritical
point would appear to change the value 
of the linear damping for the $2\acs{\tilde \omega}/2$ tongue, this is
in fact not the dominant effect.  This can be seen in 
figure~\ref{fig:lscandFM}(c) and also in (f) where 
the Floquet multipliers are plotted for both $\acs{\tilde a_5}=0$ and
a value of $\acs{\tilde a_5}$ near the bicritical point. It can be seen that 
increasing the amplitude of
the $j=5$ mode in the forcing promotes the $5\acs{\tilde \omega}/2$ tongue, as
expected, but has little impact on the damping of the other
modes.  In contrast, now comparing
figure~\ref{fig:lscandFM}(c) with (f), we see that  
the addition of the $j=2$ mode does have the effect of  
reducing the damping of the mode 
associated with the $2\acs{\tilde \omega}/2$ tongue as seen by an increase
in the Floquet Multiplier from approximately $0.5$ for $\acs{\tilde a_2}=0$ 
to $0.9$ for $\acs{\tilde a_2}=0.8\acs{\tilde g}$. 

Rather than changing the damping of the weakly damped mode, 
increasing the amplitude of the $j=5$ mode, so heading towards
the bicritical point, has the effect of increasing the quadratic
coefficients $\alpha_1$ and $\alpha_2$:  
in the limit when $\acs{\tilde a_5}=0$ the quadratic
coefficients are zero.  This is because
the temporal resonance condition 
discussed in section \ref{subsect:theoreticalideas} is not met 
when $\acs{\tilde a_5}=0$, so there can be no coupling between
the $2\acs{\tilde \omega}/2$ mode and the $4\acs{\tilde \omega}/2$ mode.

\begin{figure}
\begin{center}
\includegraphics[scale=1.0]{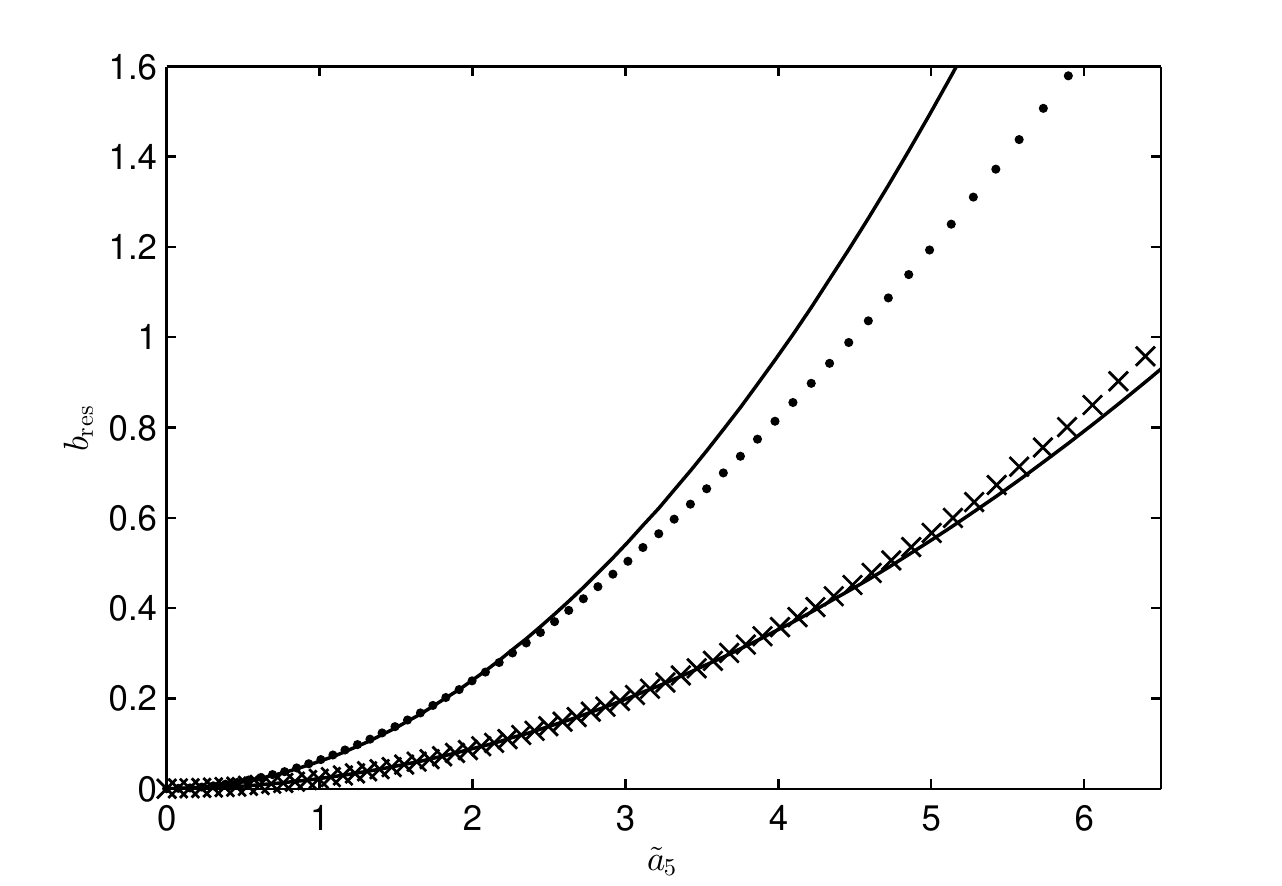}
\end{center}
\caption{The value of $b_{res}$ for $j=\{4,5\}$ (crosses) and for $j=\{4,5,2\}$
(dots) for increasing $\acs{\tilde a_5}$.  
The solid lines are proportional to $\acs{\tilde a_5}^2$.
\acs{Parameter
values as in \cite{DU_PRE06} but with $\tilde \nu=0.21$cm$^2$s$^{-1}$.} 
}
\label{fig:bres}
\end{figure}

\cite{TPS_PRE04} consider a weak viscosity limit and show that in the case
of $j=\{m,n\}$ forcing the temporal resonance condition results in possible
three-wave interactions with an $(n-m)$ mode, where one expects 
\begin{equation}
b_{\rm res} = \frac{\alpha |a_n|^2}{|\lambda_2|},
\label{eq:Porterbres}
\end{equation}
where $\lambda_2$ is the linear damping of the weakly 
damped mode and $\alpha$  is a coefficient.  
Although this cannot necessarily be expected to apply to the moderate
viscosity case of the experiments in \cite{DU_PRE06}, we see
in figure~\ref{fig:bres} that the calculated values for 
$b_{res}$ for the $j=\{4,5\}$ case have a close to quadratic
dependence on $a_5$ as suggested by equation (\ref{eq:Porterbres}).
For $j=\{m,n,p\}$ 
frequency forcing that has a three-wave resonance
with an $(n-m)$ mode, as for the case $j=\{4,5,2\}$,
\cite{TPS_PRE04} find that the dominant contribution to 
$b_{\rm res}$ is given by
\begin{equation}
b_{\rm res} = \alpha |a_n|^2 P_2(\Phi),
\label{eq:Porterbres_mnp}
\end{equation}
where $\Phi=\phi_p+2\phi_m-2\phi_n$ and $P_2$ is given by
$$
P_2 = \frac{|\lambda_2| + \mu a_2 \sin \Phi}{
            |\lambda_2|^2 - \mu^2 a_2^2},
$$
where $\mu$ is a coefficient. As $a_2 \to 0$, 
equation (\ref{eq:Porterbres_mnp}) reduces to 
equation (\ref{eq:Porterbres}). However, for
$a_2$ non-zero, the effect of $P_2$ is 
to make the leading order dependence 
of $b_{\rm res}$ on $a_2$ not purely quadratic, so
it is not surprising that for this case, 
an assumption of quadratic dependence fits less
well.  

Note that the $\Phi$ dependence,
which comes from a parameter symmetry,
does explain the strong diagonal structure to
the figures showing the pattern dependence as a function
of two of the phases, see figures~\ref{fig:DU_fig4b}
and \ref{fig:DU_fig6b}.

\begin{figure}
\begin{center}
\includegraphics[scale=1.0]{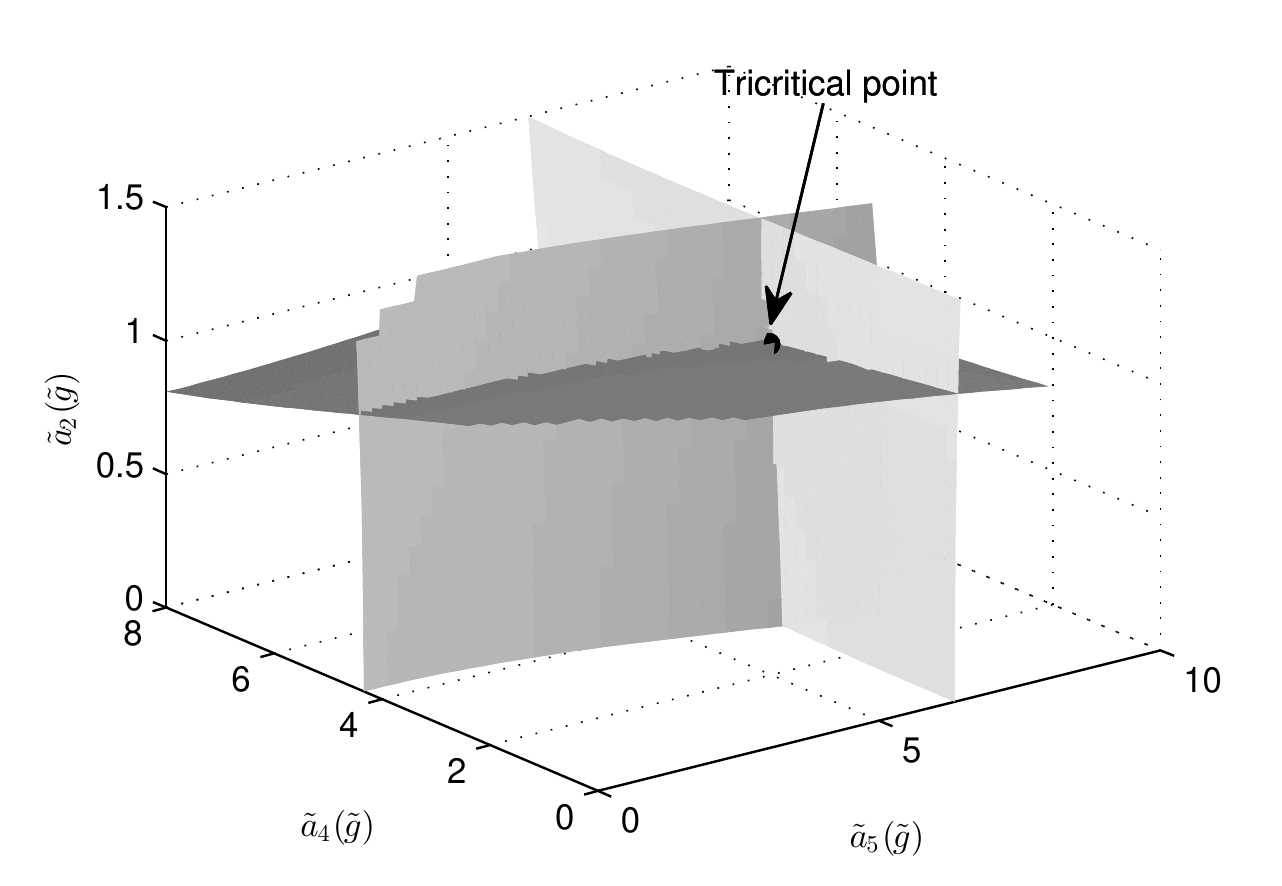}
\end{center}
\caption{Surfaces for the onset of modes with frequencies $4 \acs{\tilde \omega}/2, 
5 \acs{\tilde \omega}/2$ and $2\acs{\tilde \omega}/2$. \acs{Parameter
values as in \cite{DU_PRE06} but with $\tilde \nu=0.21$cm$^2$s$^{-1}$.} 
}
\label{fig:closenesstocod3}
\end{figure}

There is a further issue to consider.
Increasing the amplitude of the $j=2$ mode brings
the location of the bicritical point where
both the $4\acs{\tilde \omega}/2$ and the $2\acs{\tilde \omega}/2$ mode onset
simultaneously closer to the pattern onset point.  This
is seen in figure~\ref{fig:DU_fig3b} which shows both the
linear stability boundaries for both $4\acs{\tilde \omega}/2$ and
$2\acs{\tilde \omega}/2$ modes.  The proximity of the linear stability
boundary means that, 
in order to quantitatively predict the regions
of quasipatterns, it is likely that one
would need to consider the tri-critical problem
where $4\acs{\tilde \omega}/2, 5\acs{\tilde \omega}/2$ and $2\acs{\tilde \omega}/2$ modes all occur
simultaneously.
The onset boundaries for the three modes for
$j=\{4,5,2\}$ are shown in figure~\ref{fig:closenesstocod3}:
the position of the tri-critical point
is at approximately $(\acs{\tilde a_4,\tilde a_5,\tilde a_2})=(3.5,6.4,1.0)\acs{\tilde g}$.
The situation for $j=\{6,7,2\}$ forcing is similar, and
one explanation for discrepancies between the weakly nonlinear theory 
and experimental results may be that they are a consequence of the proximity of the 
codimension three point.  This kind of discrepancy was previously
demonstrated by \cite{JR_THESIS12} in a comparison of a codimension one and a codimension
two analysis of superlattice patterns in a model PDE.

Another important feature of the analysis considered here
is that, the codimension one description of the problem
results in amplitude equations 
that are variational.  This variational structure is
useful and it enables us
to construct a Lyapunov functional to predict the `most stable' 
pattern. 
However, this variational framework also means that all
predicted patterns repeat with either the period or
half the period of the drive 
but cannot exhibit more complicated time dependence.
As discussed in \cite{RSS_PRL12}, in the vicinity of the
bicritical point and depending on the signs of the
quadratic terms, an infinite sequence of resonances
can take place leading to spatio-temporal chaos. This 
may explain some of the disordered states seen close to
onset in the experiments in \cite{EF_PRL04} and 
\cite{DU_PRE06}.
The difficulties of analysing and numerically solving 
the time-dependent Navier-Stokes equations leave\acs{\sout{s}} a clear role for 
reduced models here.
For example, the model of \cite{ZV_JFM97} is derived from the
Navier--Stokes equations under the 
assumptions of infinite depth and weak viscosity but, as was
shown in \cite{SP_PRE11}, weakly nonlinear analysis of the Zhang--Vi{\~n}als
equations agrees surprisingly well with weakly nonlinear
analysis of the full Navier--Stokes equations even for the moderate 
viscosity values used in many Faraday experiments.

Finally, our results suggest that weakly nonlinear theory 
is a useful tool to understand the pattern transitions near 
onset, but there are some caveats.
Ideally experiments and theory would be done hand-in-hand
otherwise it may be hard to get agreement even in the linear
theory, as we saw in section 3;  or establish which modes
are present in patterns, as we saw in section 4; 
or establish if there
is bistability, as we saw in section 5. There are even 
problems at the basic level of the definition of the
forcing in equation (\ref{NSnd}).
In single frequency experiments changing the sign of $f(t)$ is 
equivalent to a phase shift in time, so the sign of the forcing is not 
important.
However, for multiple frequencies the analogous result is perhaps less 
transparent: changing the sign requires not just a translation
in time but requires an altered definition for the phases.
The particular sign convention used by any one experimental group
will depend on how the accelerometer
is wired up to the experiment.  Specifically, it will depend
on whether or not the accelerometer records its maximum value when
the oscillator is at its highest point or at its lowest.
This level of detail is not normally recorded ---
perhaps because
an assumption is made that it does not matter.  
However, this
can lead to some confusion: for example, the phase invariant $\Phi$ discussed in
\cite{PTS_PRL04} is derived with one particular choice of sign for 
$f(t)$ and thus one specific definition of the phases for the forcing
components.  \cite{DU_PRE06} note that they find a 
difference between the position of the maximum value of the phase 
invariant and the theory of \cite{PTS_PRL04} and between their results
and those of \cite{EF_PRE06}, but in both cases these differences can 
be explained by a different convention for the sign of the forcing 
$f(t)$.  
\acs{Based on the observations in \cite{DU_PRE06}, we have used the 
same sign convention as \cite{PTS_PRL04} for our comparison with the
work of \cite{EF_PRE06} and the opposite sign convention for \cite{DU_PRE06}.}


\appendix

\section{Calculating Floquet multipliers} \label{appA}

The Floquet multipliers can be calculated by considering 
the linear stability of the
trivial solution (\ref{eq:trivialsol}) of equation
(\ref{NSndf}) and associated
boundary conditions to perturbations of the divergence free form 
$$
{\bf u} = 
\left ( \frac{i}{k} \partial_z W(z,t)  ,0,  W(z,t) \right ){\rm e}^{ikx}.  
$$
This leads to a fourth order equation for $W(z,t)$ for $-\acs{\tilde h}/\acs{\tilde l}<z<0$
\begin{equation}
\left ( \partial_t - C \left ( - k^2 + \partial_{zz} \right ) \right )
       \left ( -k^2 + \partial_{zz} \right ) W = 0,
\label{eq:4W}
\end{equation}
with boundary conditions at $z=-\acs{\tilde h}/\acs{\tilde l}$,
$$
W(-\acs{\tilde h}/\acs{\tilde l},t) = \partial_z W|_{-\acs{\tilde h}/\acs{\tilde l}} = 0
$$
and at $z=0$,
\begin{eqnarray*}
\partial_t Z & = & W, \\
\left ( k^2 + \partial_{zz} \right ) W  & = & 0, \\
\left ( \partial_t + C \left ( 3 k^2 - \partial_{zz} \right ) \right ) \partial_z W
& = & - \left ( (1 + \acs{\cancel{a}} f(t)) k^2 + B k^4 \right ) Z. 
\end{eqnarray*}
A finite difference discretization of equation (\ref{eq:4W}) and
its boundary conditions is then carried out by letting $W_j^n = W(z_j, t_n),
j=0..J, z_j = -\acs{\tilde h}/\acs{\tilde l}+\acs{\tilde h}j/\acs{\tilde l}J$ and 
$t_n = n \delta t$
and $Z^n = Z(t_n)$. The resulting map is of the general form
\begin{equation}
{\bf A}                 \left ( \begin{array}{c} 
                       W_1^{n+1} \\
                       W_2^{n+1} \\
		       \vdots \\
		       W_{J-1}^{n+1}   \\
		       W_J^{n+1}   \\
		       Z^{n+1}
		       \end{array} \right )  =   
{{\bf B} ( t_n) } \left ( \begin{array}{c} 
                       W_1^n \\
                       W_2^n \\
		       \vdots \\
		       W_{J-1}^n  \\
		       W_J^n  \\
		       Z^n \end{array} \right ), 
\label{eq:map}
\end{equation}
where {\bf A} and ${\bf B}$ are $(J+1)\times(J+1)$ matrices
given by
$$
{\bf A} = \left ( \begin{array}{c  c c c c c c}
\acs{r}
  & 1
    & 0
       & 
         & 
	   & 
	     & \\
1
  & \acs{r}
  & 1
       & 0
         & 
	   & 
	     & \\
0 
  & 1
    & \acs{r}
      & 1
        & 0
          & 
	    & \\
\vdots 
  & 
    & 
      & \ddots
        & 
	  & 
	    & \vdots \\
& 
  & 0
     & 1
       & \acs{r}
         & 1 
	    & 0\\
  & 
    & 
      & 0
	& 2
          & \acs{r}
            & 0\\
  & 
    & 
      & 
	& 
          & 0 
	    & 1
\end{array} \right )
$$
and
$$
{\bf B} = \left ( \begin{array}{ccccccccc}
c + C \delta t /\delta z^2
  & b
    &  C \delta t/ \delta z^2 
      & 0
        &  
	  & 
	    &
	      & 
	        &  \\
b 
  & c
    &  b 
      & C \delta t/ \delta z^2
        & 0
	  & 
	    &  
	      & 
	        &  \\
C \delta t/\delta z^2
  & b 
    & c
      &  b 
        & C \delta t/ \delta z^2
          & 0
	    &  
	      & 
	        &  \\
0
  & C \delta t/\delta z^2
    & b 
      & c
        &  b 
          & C \delta t/ \delta z^2
            & 0 
	      & 
	        &  \\
\vdots
  & 
    & 
      & 
        &  \ddots
          & 
            & 
	      & 
	        &  \vdots \\
  & 
    & 0 
      & C \delta t/\delta z^2
        & b 
          & c 
            & b
	      & C \delta t/\delta z^2 
	        &  0 \\
  &
    & 
      & 
        & C \delta t/\delta z^2
          & b 
            & c - C\delta t/\delta z^2 
              & b +d 
	        & 0 \\
  &
    & 
      & 
        &
          & 2C \delta t/\delta z^2
            & b + e 
              & c +g
                & {\tilde f}^n \\
  &
    & 
      & 
        &
          &  
            & 0
              & \delta t
                & 1 \\
\end{array} \right ).
$$
where
\begin{eqnarray*}
\acs{r} & = & - \left ( 2 + k^2 \delta z^2 \right ) \\
b & = & 1 - 2Ck^2 \delta t - 4C \frac{\delta t}{\delta z^2} \\
c & = &  -\left (2 + k^2 \delta \acs{\cancel{x}z}^2 \right ) + C \frac{\delta t}{\delta z^2} 
 \left (6 + 4 k^2 \delta z^2 + k^4 \delta z^4 \right ) \\ 
d & = &  C \frac{\delta t}{\delta z^2} \left ( 2 - k^2 \delta z^2 \right )\\
e & = & 1 - 4Ck^2 \acs{\delta t} \\
{f}^n & = &  2 \delta t \delta z
  \left ( \left ( 1 + f(t_n) \right ) k^2 + Bk^4 \right ) \\ 
g & = & -C \frac{\delta t}{ \delta z^2} 
  \left (2 - k^2 \delta z^2 \right )^2.
\end{eqnarray*}
The matrix ${\bf B}$ contains time dependence through the term
${f}^n (t_n)$.

The map (\ref{eq:map})
is iterated through one period, $T$, of the drive 
resulting in a map ${\bf W}^N = {\bf D W}^0$  
that takes
${\bf W}^0$ to ${\bf W}^N$ where $N=T/\delta t$. 
The eigenvalues of the matrix $\bf D$ then give the required Floquet multipliers.

\vspace{1cm}

This paper builds on discussions about pattern formation in the Faraday
problem with many individuals over the years. In 
particular the authors would like to acknowledge the input of 
Mary Silber, Jeff Porter and Jay Fineberg. \acs{We thank the referees for
their thorough reading of the paper and their helpful comments.}

\bibliographystyle{jfm}

\input{jfm_RSS_revised.bbl}
\end{document}

%% file: fig1.tex
%
%
%
%

\centering
\makebox[\hsize]{%
\mbox{\beginpgfgraphicnamed{rss_fig_geom}%
\begin{tikzpicture}[scale=1.00,>=stealth]
\path[use as bounding box] (0,0) rectangle (11,6);
  \draw[samples=50,ultra thick,domain=2:9] plot (\x,{2.5+0.5*sin((\x-2)*360*2.5/7)});
  \draw[thick,dashed] (9,2.5) -- (2,2.5) node[anchor=east] {$z=0$};
  \draw[thick] (2,3.5) -- (2,1) -- (9,1) -- (9,3.5);
  \node[anchor=east] at (2,1) {$z=-\acs{\tilde h/\tilde l}$};
  \draw[thick,<-] (6,3) -- (7,4) node[anchor=west] {$z=\zeta(x,y;t)$};
  \draw[ultra thick,->] (10,3) -- (10,2);
  \node[anchor=west] at (10.1,2.5) {$\acs{\tilde g}$};
  \draw[ultra thick,<->] (5.5,0.8) -- (5.5,0);
  \node[anchor=west] at (5.6,0.4) {$f(t)$};

\end{tikzpicture}\endpgfgraphicnamed}

}

%% file: fig3.tex
\begin{center}
\setlength{\unitlength}{1cm}
\begin{picture}(13,15)(-1,0)
\put(0,9){\resizebox{4.5cm}{!} {\includegraphics{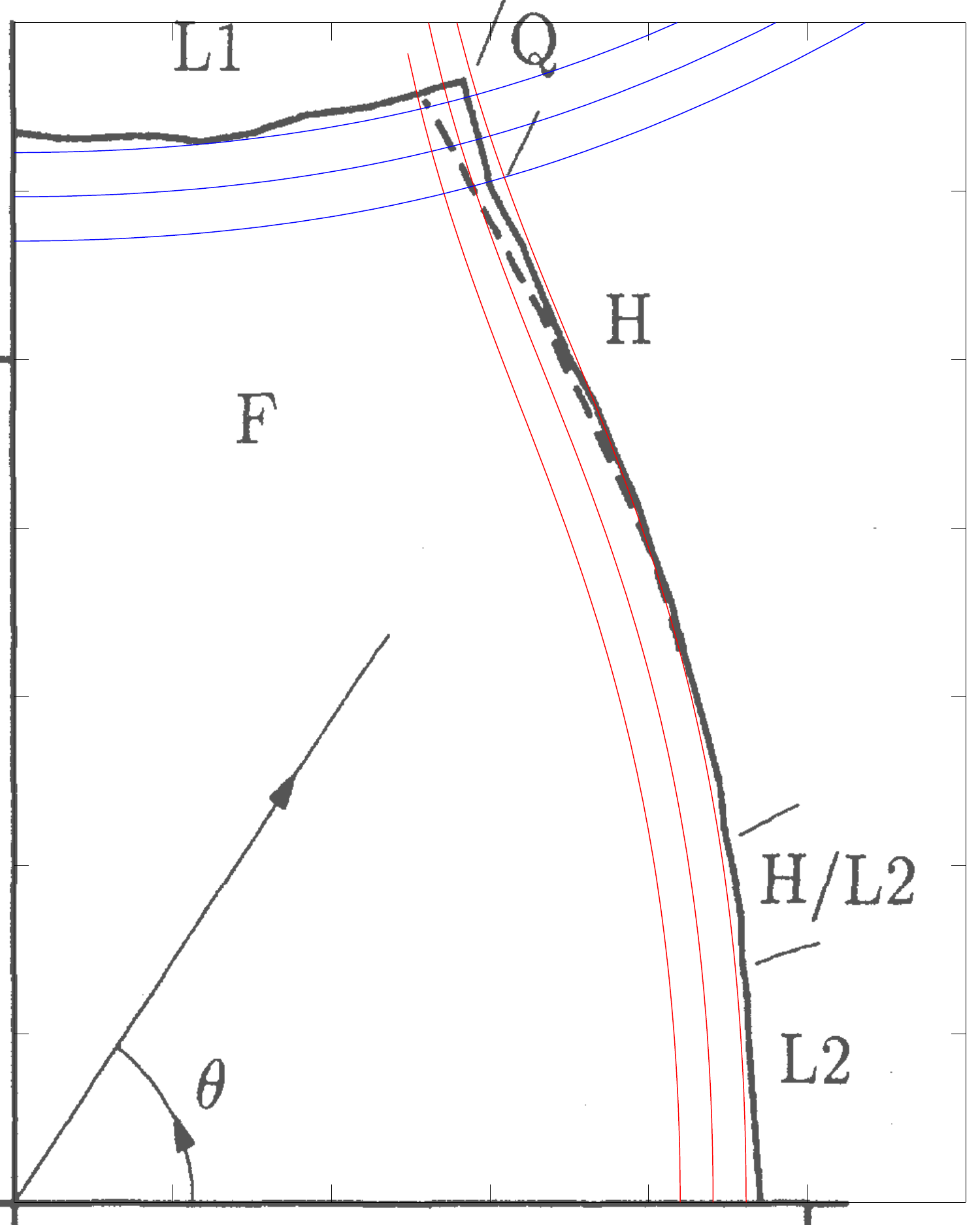}}}
\put(5.5,9){\resizebox{4.5cm}{!} {\includegraphics{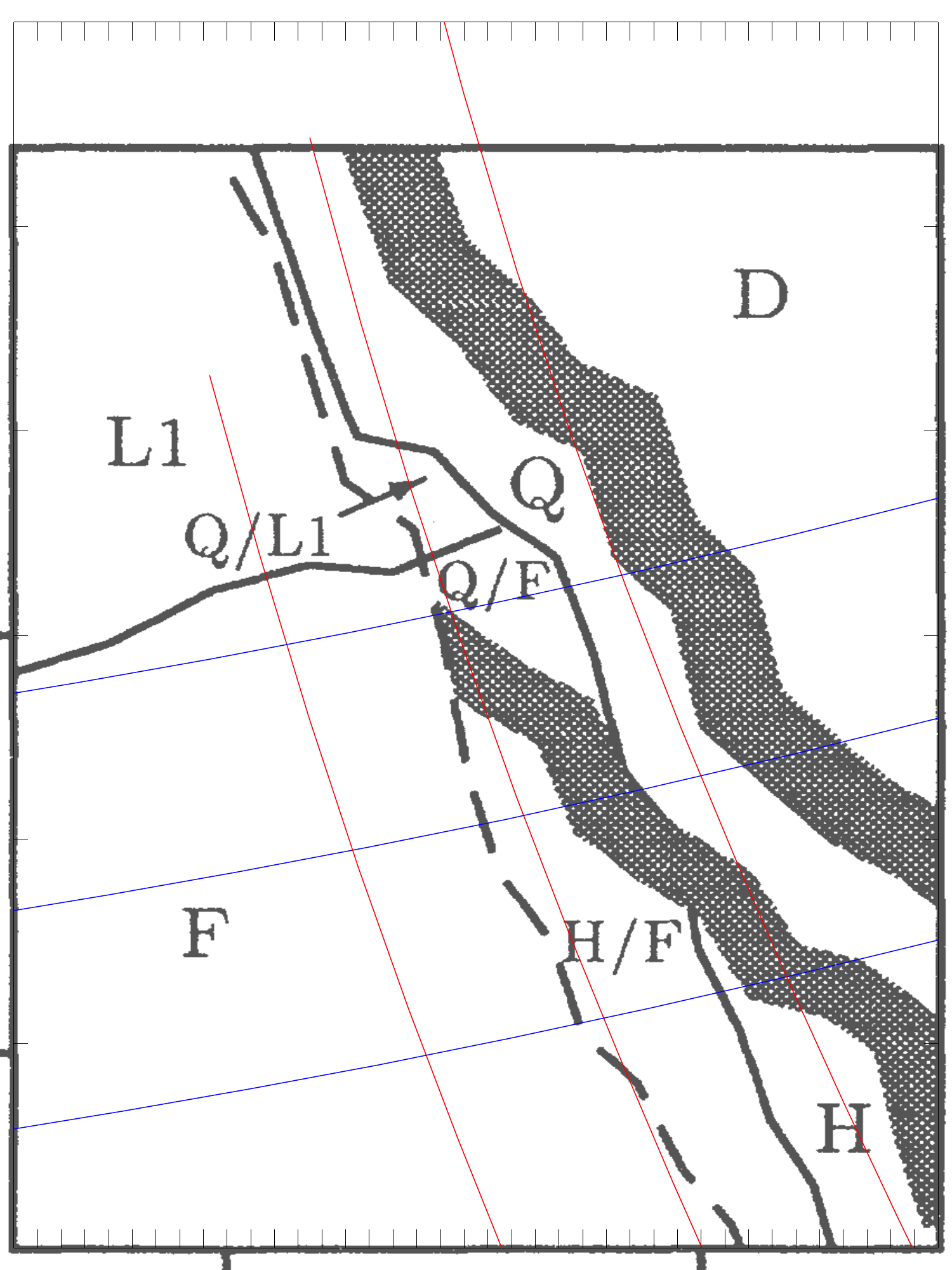}}}
\put(0,2.8){\resizebox{4.5cm}{!} {\includegraphics{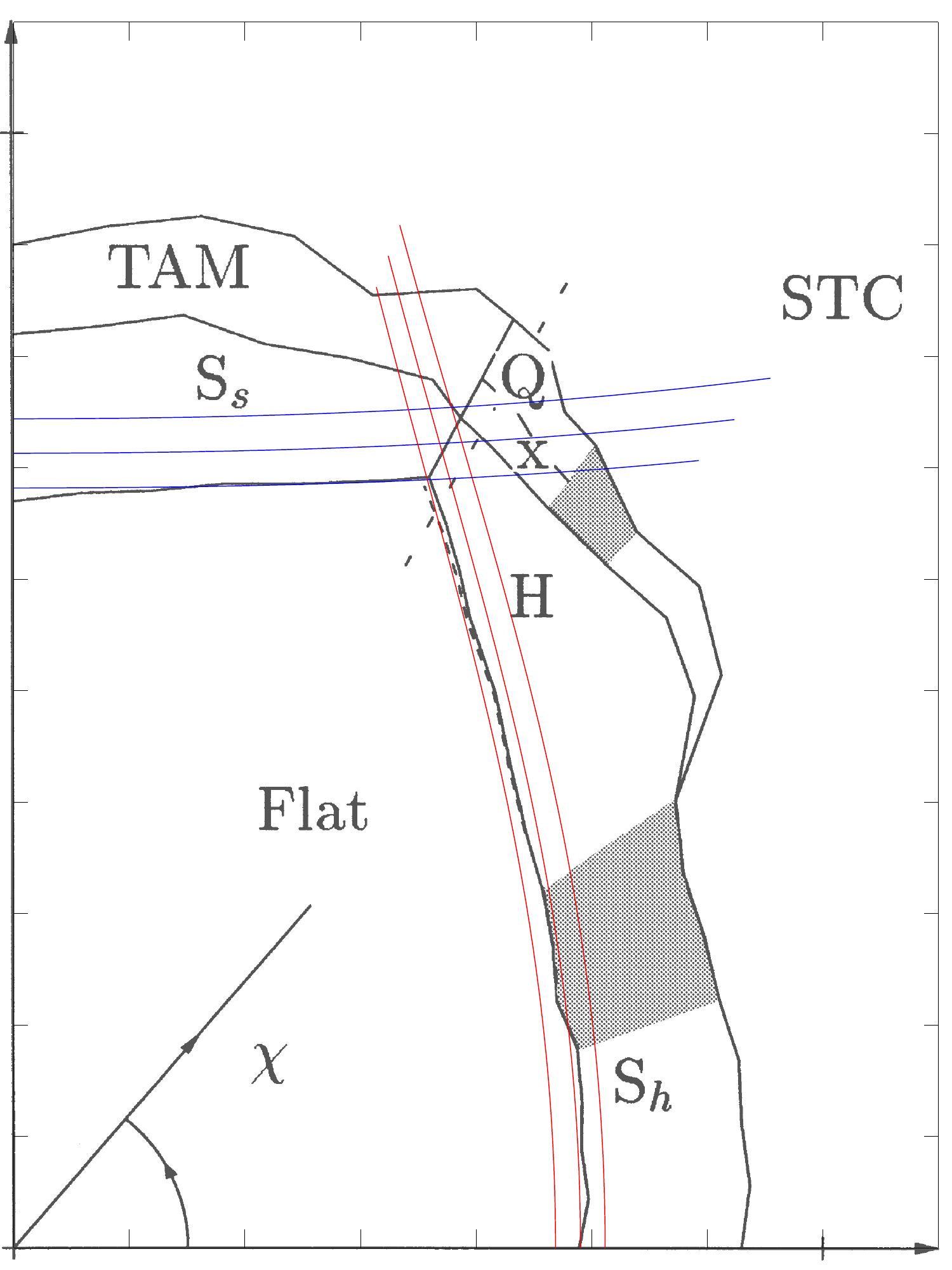}}}
\put(5.5,4.8){\resizebox{4.5cm}{!} {\includegraphics{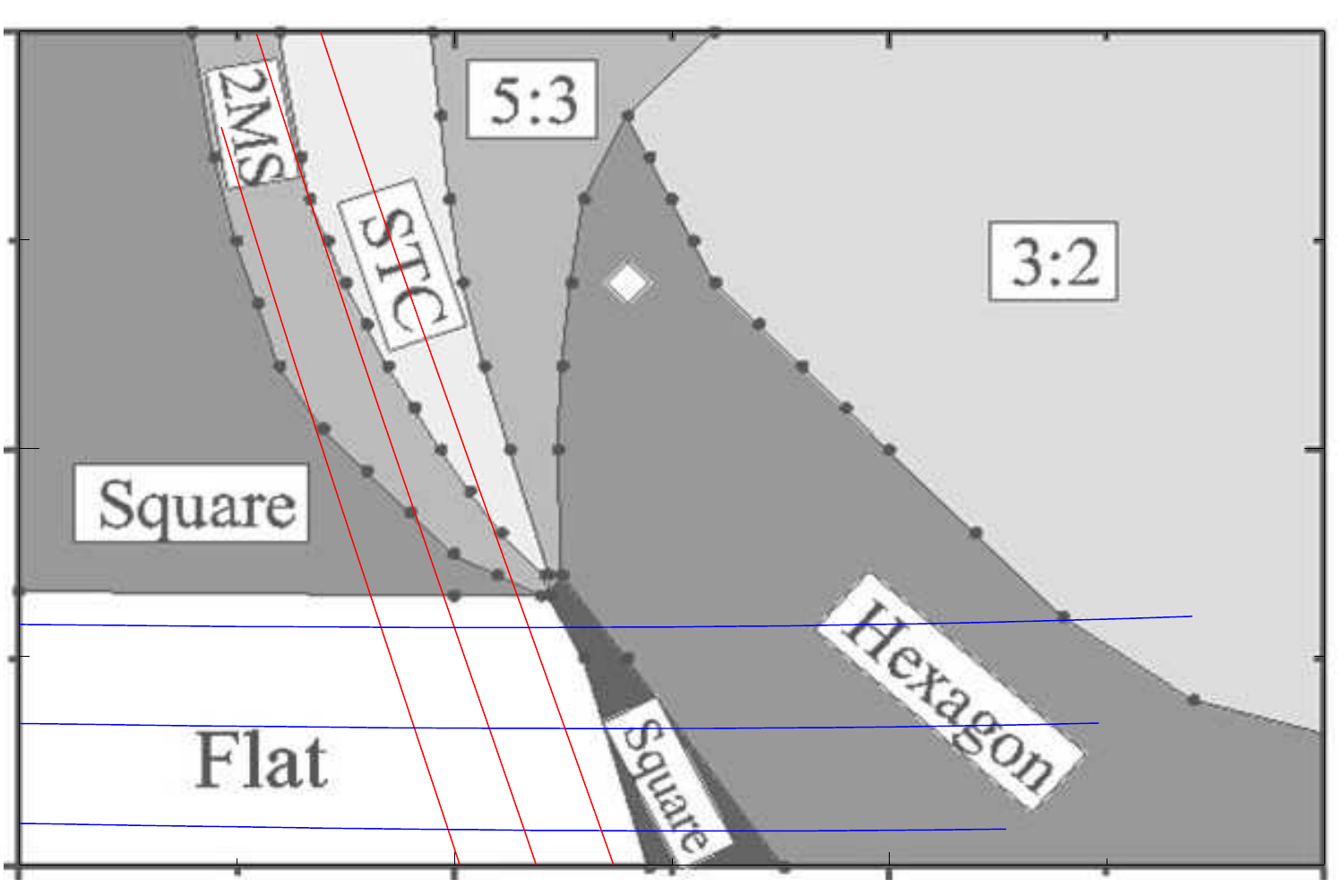}}}

\put(0,0){\resizebox{4.5cm}{!} {\includegraphics{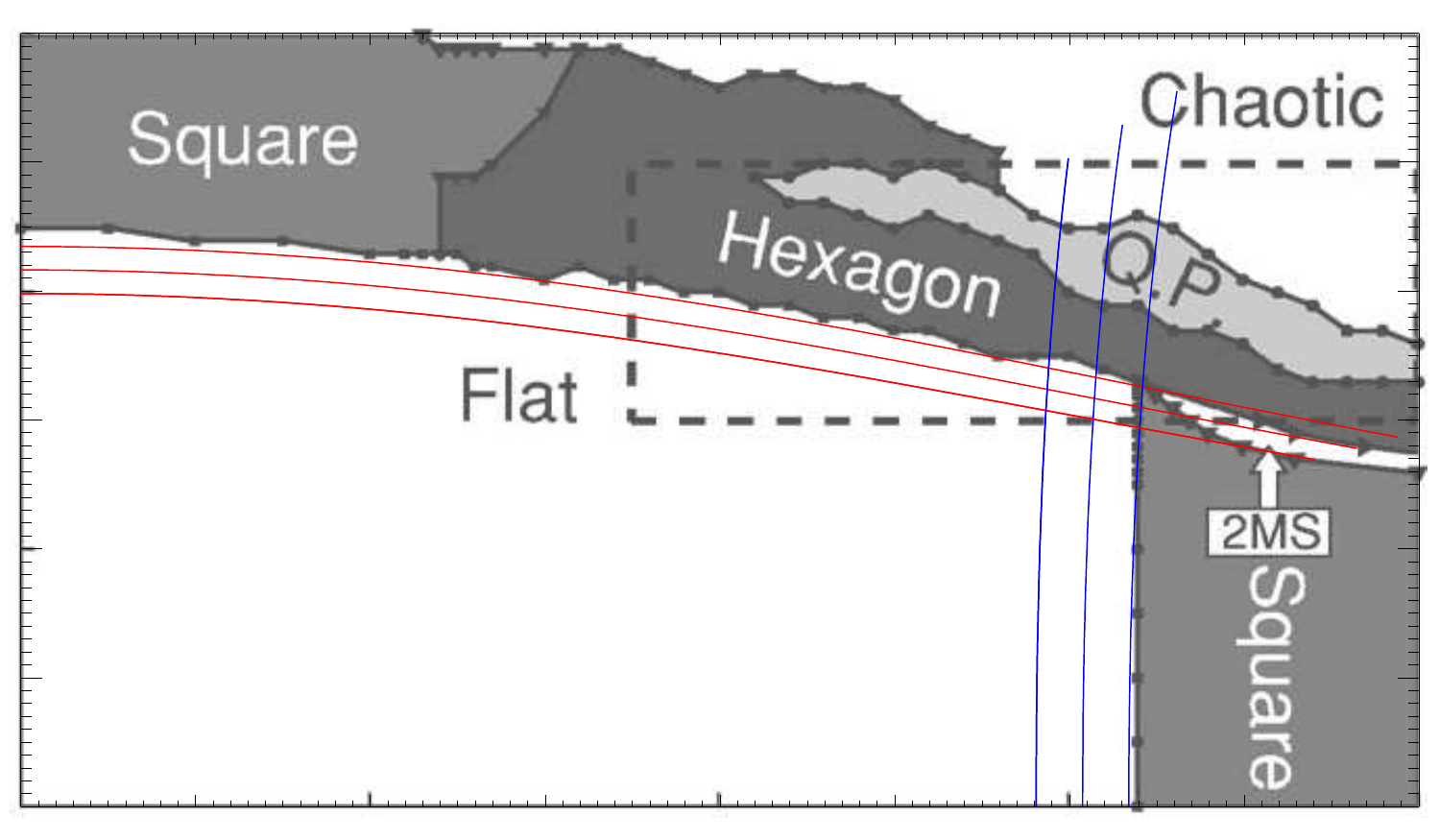}}}
\put(5.5,0){\resizebox{4.5cm}{!} {\includegraphics{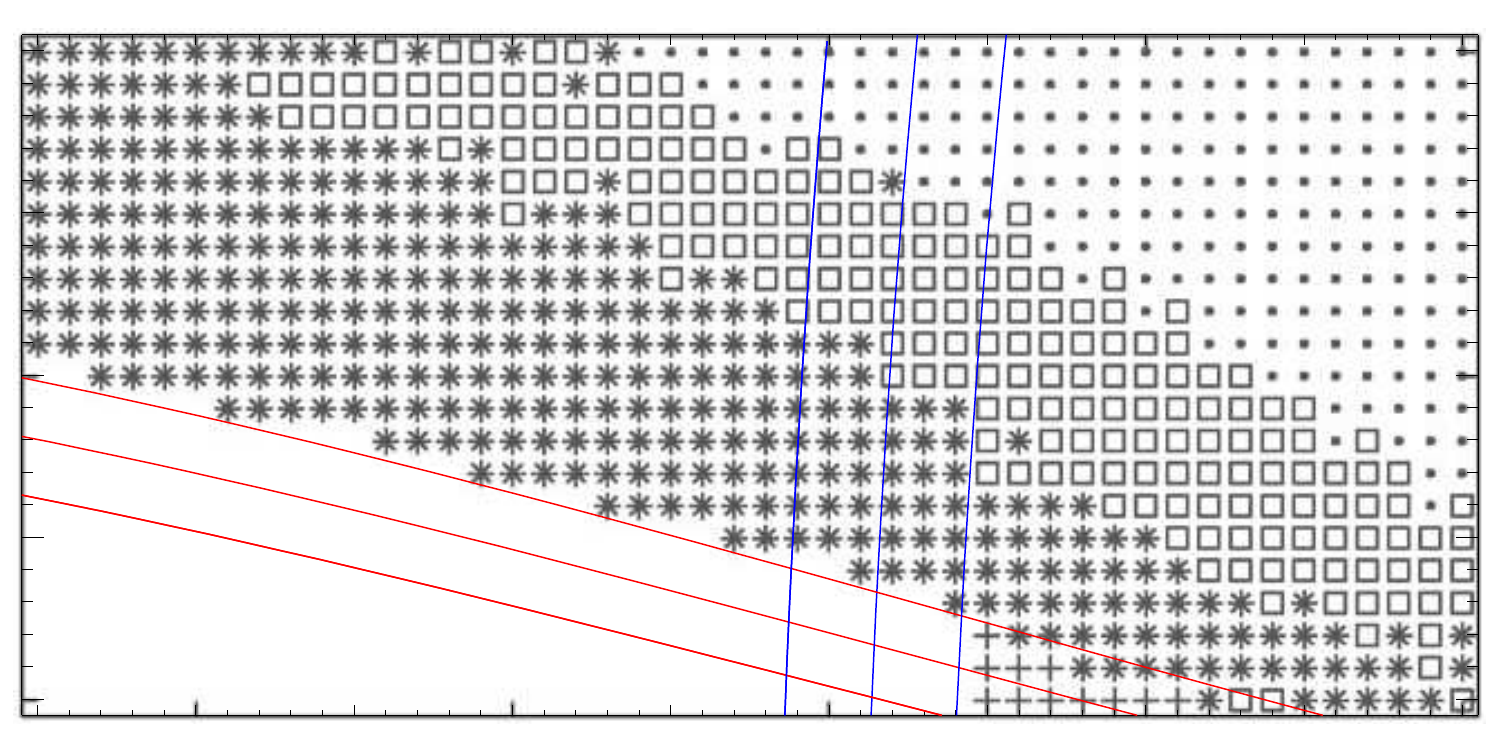}}}

\put(-0.7,14.5){(a)}
\put(5,14.5){(b)}
\put(-0.7,8.5){(c)}
\put(5,8){(d)}
\put(-0.7,2){(e)}
\put(5,2){(f)}

\end{picture}
\end{center}

%% file: fig4.tex
%
%
%
%

\makebox[\hsize]{%
\mbox{\beginpgfgraphicnamed{rss_fig_threewaves_k60}%
\begin{tikzpicture}[scale=1.00,>=stealth]
\path[use as bounding box] (-1.2,-2.0) rectangle (1.2,2.0);
\draw (-1,1.1) node[above] {(a)};
\draw[thick] (0,0) circle (1.0);
\draw[thick,->] (0,0) -- (  0.8660, -0.500) ; 
\draw[thick,->] (0,0) -- ( -0.8660,  0.500) node[above,left] {$\bk_2$};
\draw[thick,->] (0,0) -- (  0.8660,  0.500) node[above,right] {$\bk_1$};
\draw[thick,->] (0,0) -- ( -0.8660, -0.500) ; 
\draw[thick,->] (0,0) -- (  0.8660,  0.500) ;
\draw[thick,->] (0,0) -- (  0.0,     1.000) node[above] {$\bk_3$};;
\draw[thick,->] (0,0) -- (  0.0,    -1.000) ;
\draw [domain=-30:30] plot ({0.65*cos(\x)} ,{0.65*sin(\x)});
\draw (0.1,0.0) node[above=0pt, right=0pt] {$60^\circ$};
\end{tikzpicture}\endpgfgraphicnamed}
\hfil
\mbox{\beginpgfgraphicnamed{rss_fig_threewaves_kin}%
\begin{tikzpicture}[scale=1.00,>=stealth]
\path[use as bounding box] (-1.2,-2.0) rectangle (1.2,2.0);
\draw (-1,1.1) node[above] {(b)};
\draw[thick] (0,0) circle (1.0);
\draw[thick,dashed] (0,0) circle (0.6600);
\draw[thick,->] (0,0) -- (  0.3300,  0.9440) node[above=6pt,right=0pt] {$\bk_2$};
\draw[thick,->] (0,0) -- ( -0.3300,  0.9440);
\draw[thick,->] (0,0) -- ( -0.3300, -0.9440) ;
\draw[thick,->] (0,0) -- (  0.3300, -0.9440) node[below=6pt,right=0pt] {$\bk_1$};
\draw[thick,->] (0,0) -- (  0.6600,  0.0000) node[above=0pt,right=0.0pt] {$\bk_3$};
\draw[thick,->] (0,0) -- ( -0.6600,  0.0000);
\draw [domain=251:289] plot ({0.5*cos(\x)} ,{0.5*sin(\x)});
\draw (0.0,-0.3) node[above=0pt, right=-6pt] {$\theta$};
\end{tikzpicture}\endpgfgraphicnamed}
\hfil%
\mbox{\beginpgfgraphicnamed{rss_fig_threewaves_kout}%
\begin{tikzpicture}[scale=1.00,>=stealth]
\path[use as bounding box] (-1.8,-2.0) rectangle (2.2,2.0);
\draw (-1.5,1.1) node[above] {(c)};
\draw[thick] (0,0) circle (1.0);
\draw[thick,dashed] (0,0) circle (1.6400);
\draw[thick,->] (0,0) -- (  0.8190, -0.5740) node[above=0pt,right=0pt] {$\bk_1$};
\draw[thick,->] (0,0) -- ( -0.8190,  0.5740);
\draw[thick,->] (0,0) -- (  0.8190,  0.5740) node[above=0pt,right=0pt] {$\bk_2$};
\draw[thick,->] (0,0) -- ( -0.8190, -0.5740);
\draw[thick,->] (0,0) -- (  1.64, 0.0) node[above=0pt,right=0pt] {$\bk_3$};
\draw[thick,->] (0,0) -- ( -1.64, 0.0) ;
\draw [domain=-35:35] plot ({0.5*cos(\x)} ,{0.5*sin(\x)});
\draw (0.2,0) node[above=0pt, right=0pt] {$\theta$};
\end{tikzpicture}\endpgfgraphicnamed}%

\mbox{\beginpgfgraphicnamed{rss_fig_threewaves_threecircles}%
\begin{tikzpicture}[scale=1.00,>=stealth]
\path[use as bounding box] (-1.8,-2.0) rectangle (2.2,2.0);
\draw (-1.5,1.1) node[above] {(d)};
\draw[thick] (0,0) circle (1.0);
\draw[thick,dashed] (0,0) circle (0.6400);
\draw[thick,dashed] (0,0) circle (1.3800);
\draw[thick,->] (0,0) -- (  0.9039, -0.4277) node[above=0pt,right=0pt] {$\bk_1$};
\draw[thick,->] (0,0) -- ( -0.9039,  0.4277);
\draw[thick,->] (0,0) -- (  0.4760,  0.4277) node[above=0pt,right=0pt] {$\bk_2$};
\draw[thick,->] (0,0) -- ( -0.4760, -0.4277);
\draw[thick,->] (0,0) -- (  1.38, 0.0) node[above=0pt,right=0pt] {$\bk_3$};
\draw[thick,->] (0,0) -- ( -1.38, 0.0);
\end{tikzpicture}\endpgfgraphicnamed}%

}